\documentclass[fleqn,twoside]{article}

\usepackage{espcrc2mm}
\usepackage{graphicx}
\usepackage{amsmath}
\usepackage{amssymb}

\title{Isovector soft dipole mode in $^{6}$Be}

\author{%
A.S.~Fomichev\address[dub]{Flerov Laboratory of Nuclear Reactions, JINR, Dubna,
RU-141980 Russia},
V.~Chudoba\addressmark[dub]$^,$\address[opa]{Institute of Physics, Silesian
University in Opava, Bezru\v{c}ovo n\'{a}m.\ 13, 74601 Czech Republic},
I.A.~Egorova\address[bog]{Bogolyubov Laboratory of Theoretical Physics, JINR,
Dubna, RU-141980 Russia},
S.N.~Ershov\addressmark[bog],
M.S.~Golovkov\addressmark[dub],
A.V.~Gorshkov\addressmark[dub],
V.A.~Gorshkov\addressmark[dub],
L.V.~Grigorenko\addressmark[dub]$^,$\address[gsi]{GSI Helmholtzzentrum f\"{u}r
Schwerionenforschung, Planckstra{\ss}e
1, D-64291 Darmstadt, Germany}$^,$\address[kur]{Russian Research Center ``The
Kurchatov Institute'', Kurchatov sq.\ 1, RU-123182 Moscow, Russia},
G.~Kami\'nski\addressmark[dub]$^,$\address[inp]{Institute of Nuclear Physics
PAN, Radzikowskiego 152, PL-31342 Krak\'{o}w, Poland},
S.A.~Krupko\addressmark[dub],
I.G.~Mukha\addressmark[gsi],
Yu.L.~Parfenova\addressmark[dub]$^,$\address[sko]{Skobel'tsyn Institute of
Nuclear Physics, Moscow State University, 119991 Moscow, Russia},
S.I.~Sidorchuk\addressmark[dub],
R.S.~Slepnev\addressmark[dub],
L.~Standy{\l}o\addressmark[dub]$^,$\address[sol]{Andrzej So{\l}tan Institute
for Nuclear Studies, Ho\.za 69, PL-00681, Warsaw, Poland}
S.V.~Stepantsov\addressmark[dub],
G.M.~Ter-Akopian\addressmark[dub],
R.~Wolski\addressmark[dub]$^,$\addressmark[inp],
M.V.~Zhukov\address[rnfc]{Fundamental Physics, Chalmers University of
Technology, S-41296 G\"{o}teborg, Sweden}
}

\begin{document}

\begin{abstract}
By using the $^1$H($^{6}$Li,$^{6}$Be)$n$ charge-exchange reaction, continuum 
states in $^{6}$Be were populated up to $E_T=16$ MeV, $E_T$ being the $^{6}$Be 
energy above its three-body decay threshold. In kinematically complete 
measurements performed by detecting $\alpha$+$p$+$p$ coincidences, an $E_T$ 
spectrum of high statistics was obtained, containing approximately $\sim 5 
\times 10^6$ events. The spectrum provides detailed correlation information 
about the well-known $0^+$ ground state of $^{6}$Be at $E_T=1.37$ MeV and its 
$2^+$ state at $E_T=3.05$ MeV. Moreover, a broad structure extending from 4 to 
16 MeV was observed. It contains negative parity states populated by $\Delta 
L=1$ angular momentum transfer without other significant contributions. This 
structure can be interpreted as a novel phenomenon, i.e.\ the isovector soft 
dipole mode associated with the $^{6}$Li ground state. The population of this 
mode in the charge-exchange reaction is a dominant phenomenon for this reaction, 
being responsible for about $60\%$ of the cross section obtained in the measured 
energy range.

\vspace{3mm}


\vspace{1.5mm}






\end{abstract}

\maketitle


\section{Introduction}


Electromagnetic excitation is an important tool for studying light exotic (halo) 
nuclei. Based on the halo hypothesis, the existence of a novel dipole mode at 
low excitation energies was predicted 
\cite{Ikeda:1988,Hansen:1987,Bertulani:1988}. This so-called soft dipole mode 
(SDM) [or soft mode of the giant dipole resonance (GDR)] is related to the low 
binding energy of the halo nucleon(s). The low binding allows low-frequency 
oscillations of the halo nucleon(s) against the core, thus creating the 
low-lying dipole excitations and providing abnormally large cross-sections for 
electromagnetic dissociation at low-energy. Experiments confirmed these 
expectations \cite[and Refs.\ therein]{Aumann:2005} and showed that the observed 
low-lying E1 strength is in a good agreement with the cluster 
non-energy-weighted sum rule. The latter observation indicates that the SDM is 
connected with the cluster degrees of freedom in contrast with excitations in 
the GDR region, which represent collective phenomena.

The fact that $^{6}$Be is an isobaric partner of the ``classical'' halo nucleus 
$^{6}$He is of particular interest to the present work:
\begin{itemize}
\item $^{6}$Be is the lightest nucleus whose (particle unstable) ground state is 
a \emph{true two-proton} ($2p$) emitter. True $2p$ emission is a genuine 
quantum-mechanical phenomenon, in which due to the pairing effect the emission 
of one proton is energetically forbidden while the simultaneous emission of two 
protons is possible. The correlation patterns observed for the decay of $^{6}$Be 
led to the concept of \textit{democratic decay} \cite{Bochkarev:1992}  which is 
now generally accepted. In heavier $2p$-unstable nuclei the true $2p$ emission 
occurs in the form of \emph{two proton radioactivity} \cite{Goldansky:1960}, as 
discussed by comparing the correlation data of $^{6}$Be and $^{45}$Fe 
\cite{Grigorenko:2009}.
\item Considerable efforts have been made to study the halo aspect of $^{6}$He 
in the past two decades. Recently it has been demonstrated in Ref.\ 
\cite{Grigorenko:2009c} that a valuable alternative to investigating the ground 
state (g.s.) of $^{6}$He is a precision study of correlations characterizing the 
decay of the (particle-unstable) $^{6}$Be ground state.
\end{itemize}

\begin{figure}
\centerline{
\includegraphics[width=0.48\textwidth]{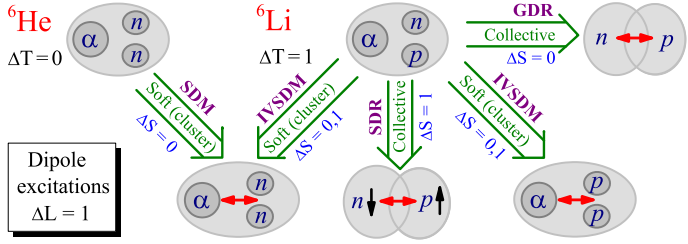}
}
\caption{Classification scheme of dipole excitations in $^{6}$He and $^{6}$Be 
produced in charge-exchange reactions with $^{6}$Li. The appearance of the soft 
dipole mode in the \emph{electromagnetic} excitation of $^{6}$He is shown for 
comparison. Given is the illustration of difference between the cluster 
excitations (modes), i.e.\ the soft dipole mode (SDM) and isovector soft dipole 
mode (IVSDM),  and the collective excitations (resonances), i.e.\ the giant 
dipole resonance (GDR) and spin-dipole resonance (SDR).}
\label{fig:mode}
\end{figure}

In the present work we studied $^{6}$Be in the charge-exchange reaction 
$^1$H($^{6}$Li,$^{6}$Be)$n$. We identified the properties of the $^{6}$Be 
continuum above the highest-lying well-established state, i.e.\ the $2^+$ state 
at $E_T=3.03$  MeV, $E_T$ being the $^{6}$Be energy above its three-body decay 
threshold. We found that in a broad energy range up to about$ E_T =16$ MeV this 
continuum contains mostly negative parity states populated by $\Delta L=1$ 
transitions. The negative-parity continuum in $^{6}$Be can be interpreted to 
present an analogy to the soft dipole mode in $^{6}$He. Based on the $\Delta 
L=1$ identification in the charge-exchange reaction, it was suggested that the 
low-energy continuum properties of $^{6}$He 
\cite{Sakuta:1993,Nakayama:2000,Nakamura:2002} and $^{6}$Be \cite{Yang:1995} may 
represent a sort of soft dipole excitation. In this work we are able to confirm 
this suggestion on the basis of the correlation data. The obtained results 
provide qualitatively new insight into this problem. We also reveal important 
differences between our results and data obtained for the SDM in $^{6}$He, which 
enables us to interpret the continuum structure of $^{6}$Be as the manifestation 
of a novel phenomenon, namely the isovector soft dipole mode (IVSDM).

The classification of the dipole excitations in $^{6}$He and $^{6}$Be built on 
$^{6}$Li g.s.\ is illustrated in Fig.\ \ref{fig:mode}. This figure allows to 
clarify the difference with the interpretation of the dipole excitations in 
Refs.\ \cite{Yang:1995,Nakayama:2000,Nakamura:2002} as a part of the GDR or SDR 
which are collective phenomena. In contrast, our interpretation of the soft 
dipole excitations in $^{6}$He and $^{6}$Be is entirely connected to the cluster 
degrees of freedom. We presume that collective resonances are excluded below the 
$\alpha$-cluster disintegration energy and that the observed spectra are thus 
connected solely with the specific three-body cluster dynamics of $A = 6$ 
nuclei. From theoretical arguments there are no low-lying $T = 1$ 
\emph{resonances} of negative parity (structures with definite energy and width) 
expected in $A = 6$ nuclei. Therefore, the strong population of the continuum 
structure in $^{6}$Be has to be interpreted in terms of \emph{excitation modes} 
which strongly depend on the excitation mechanism.

\begin{figure}
\centerline{
\includegraphics[width=0.48\textwidth]{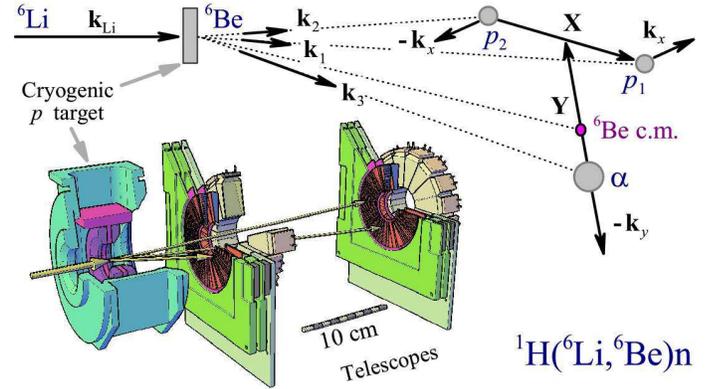}
}
\caption{Experimental setup and kinematical diagram. The coordinate space Jacobi
variables $\mathbf{X}$ and $\mathbf{Y}$ as well as the conjugated momentum space 
Jacobi variables $\mathbf{k}_x$ and $\mathbf{k}_y$ are shown for the ``T'' 
Jacobi system.}
\label{fig:setup}
\end{figure}


\section{Experiment}


A 47 A MeV $^{6}$Li beam was produced at the U-400M Cyclotron of the Flerov 
Laboratory (JINR, Dubna, Russia \cite{u400m}). The fragment separator ACCULINNA 
\cite{Rodin:1997} served as a transport line delivering the beam to the target 
position. A graphite degrader installed in front of ACCULINNA reduced the beam 
energy to 32.5 A MeV. The beam delivered to the target had an intensity of 
$3\times10^7$ s$^{-1}$, an energy spread smaller than $0.5\%$ and covered a 
circular area of 6 mm diameter.

A schematic view of the experimental setup, the kinematical diagram and the 
kinematical variables are shown in Fig.\ \ref{fig:setup}. The 6 mm thick target 
cell having 6 $\mu$m stainless-steel windows was filled with hydrogen at a 
pressure of 3 bar and cooled down to 35 K. The reaction products were detected 
by two identical annular telescopes placed 91 and 300 mm downstream of the 
target. Each telescope consisted of two position-sensitive silicon detectors of 
300 and 1000 $\mu$m thickness, respectively, and an array of 16 trapezoidal 
CsI(Tl) crystals with PIN-diode readouts. The inner and outer diameters of the 
active zone of the silicon detectors were 32 and 82 mm, respectively. The first 
detector was segmented in 32 rings on one side and 32 sectors on the other, 
whereas the second detector had 16 sectors only. The 19 mm thick CsI(Tl) 
crystals were assembled in a ring with an inner and outer diameter of 37 and 90 
mm, respectively. Angular ranges of approximately 3 to 8$^{\circ}$ and 10 to 
24$^{\circ}$ were covered by the first and the second telescopes, respectively. 
Each segment of the telescopes had its own acquisition channel. Particle 
identification was provided by the standard $\Delta E$-$E$ method.


\section{Experimental results}

\begin{figure}
\leftline{
\includegraphics[width=0.421\textwidth]{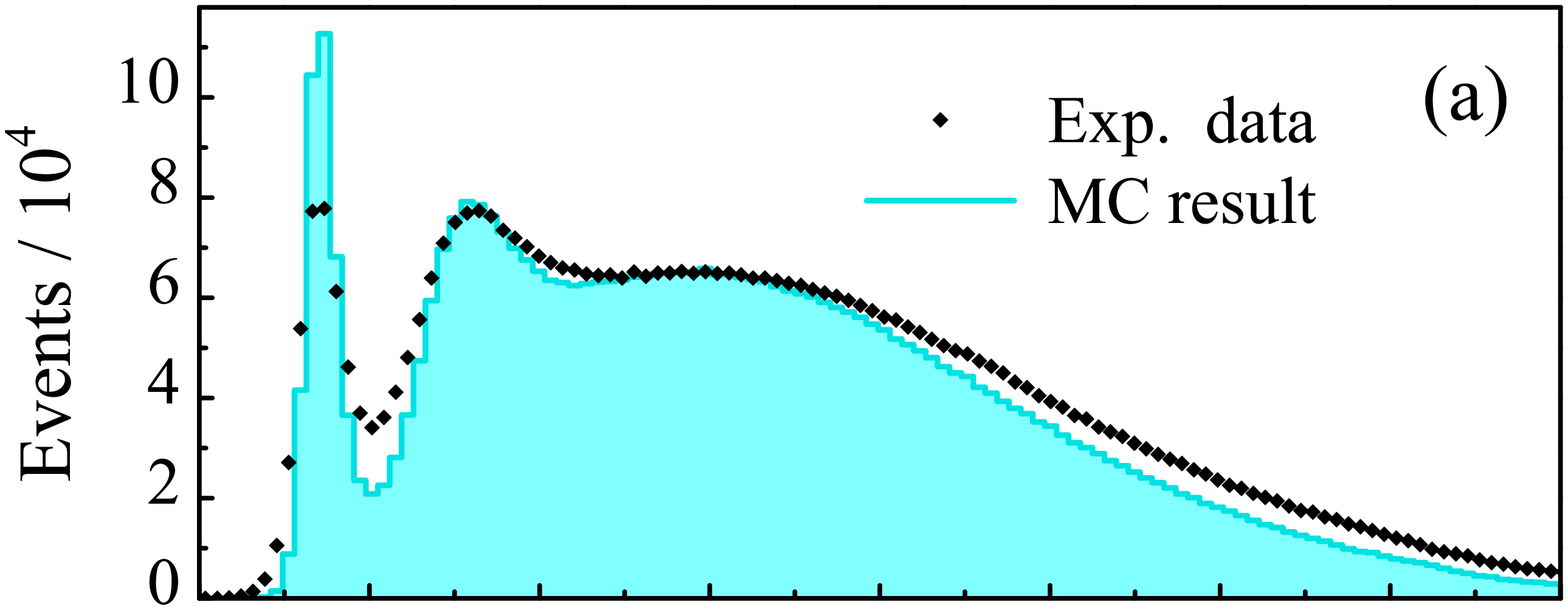}
}
\leftline{
\includegraphics[width=0.426\textwidth]{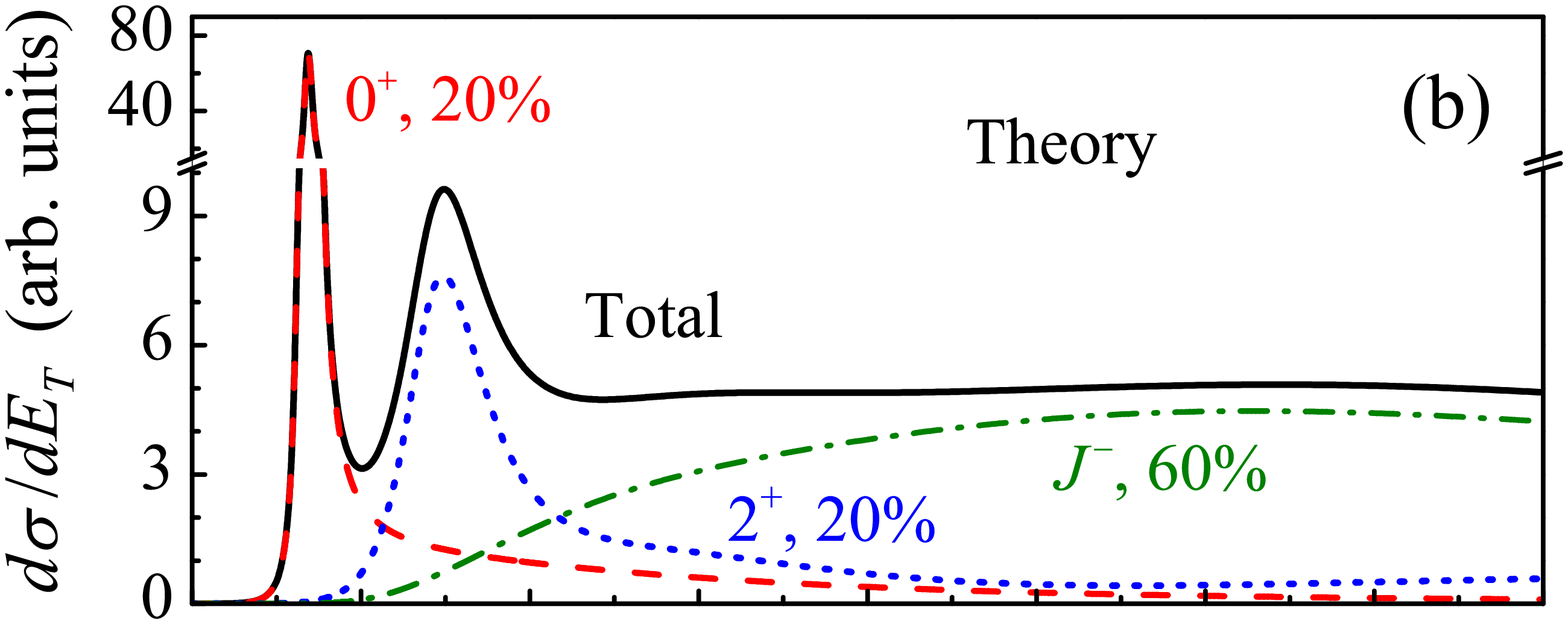}
}
\leftline{
\includegraphics[width=0.483\textwidth]{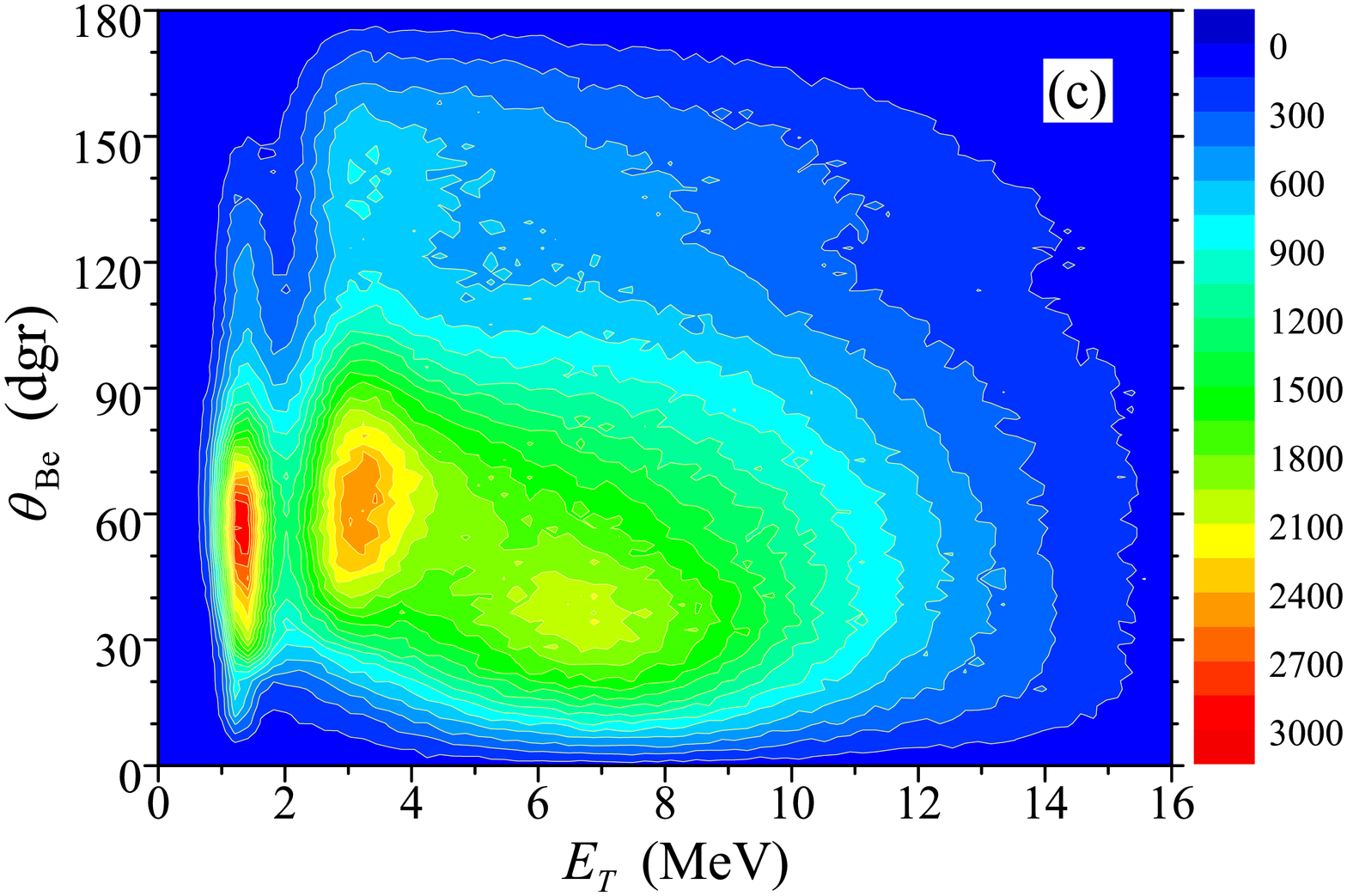}
}
\caption{(a) Experimental energy spectrum of $^{6}$Be (diamonds) and Monte Carlo 
simulation (shaded histogram). The energy bin size is 140 keV. (b) Spectrum 
obtained by correcting the experimental data for efficiency, which represents 
the input for Monte Carlo simulation; the contributions of different $J^{\pi}$ 
components is indicated. (c) Contour plot of the spectrum in the 
$\{E_T,\theta_{\text{Be}} \}$ plane.}
\label{fig:cross-sect}
\end{figure}

By analysing triple ($p$+$p$+$\alpha$) coincidence events, we determined the 
invariant mass of $^{6}$Be and its center-of-mass (c.m.) momentum vector. Due to 
the inverse kinematics of the $^{1}$H($^{6}$Li,$^{6}$Be)$n$ reaction the decay 
products  of $^{6}$Be move in forward  direction within a rather narrow angular 
cone. Therefore the two protons and the $\alpha$-particle, ejected from the 
$^{6}$Be continuum states with energies up to $E_T = 16$ MeV, were detected with 
reasonable efficiency almost in the whole angular range of 
$0^{\circ}<\theta_{\text{Be}}<180^{\circ}$. The angle $\theta_{\text{Be}}$ is 
the $^{6}$Be emission angle which is defined in the c.m.\ system of the 
$^{1}$H($^{6}$Li,$^{6}$Be)$n$ reaction. The  spectrum measured in this angular 
range is displayed in Fig.\ \ref{fig:cross-sect} (a). Due to the high 
statistics, obtained by accumulating approximately $\sim 5\times 10^6$  
$2p$-decay events, statistical experimental uncertainties are not visible in 
Fig.\ \ref{fig:cross-sect} (a). The $0^+$ g.s.\ at $E_T = 1.37$ MeV and the 
first excited $2^+$ state at $E_T = 3.05$ MeV are identified directly in this 
inclusive spectrum. In the measured spectrum, these two peaks are superimposed 
on a broad hump which starts (presumably) from the g.s.\ energy and has a 
maximum at $E_T \sim 6$ MeV. The high-energy slope of the hump is mainly 
connected to the limited angular acceptance of the detector array and to the 
decrease of the detection efficiency for protons with laboratory energies above 
45 MeV.

The width of the g.s.\ peak corresponds to the overall energy resolution of the 
experiment. The energy resolution obtained by Monte-Carlo (MC) simulations can 
be approximated to follow a $\sqrt{E_T}$ dependence, resulting in FWHM values of 
0.4, 0.6, 1.0 and 1.3 MeV for $E_T$ values of 1.4, 3, 9, and 15 MeV, 
respectively.

The two-dimensional plot $E_T$ vs.\ $\theta_{\text{Be}}$ based on the measured 
data and presented in Fig.\ \ref{fig:cross-sect} (c), reveals  that the broad 
hump extending to 16 MeV is characterized by a regular behavior in the whole 
angular range. As will be discussed in Sec.\ \ref{sec:ang-cor}, the events found 
in any part of this hump show similar regularities both in the angular 
distribution and in the correlations of the kinematical variables characterizing 
the three-body decay of $^{6}$Be.

The population of the $^{6}$Be g.s.\ in the $^{1}$H($^{6}$Li,$^{6}$Be)$n$ 
charge-exchange reaction is associated mainly with the $\Delta L = 0$ orbital 
momentum transfer, providing the cross-section maximum at zero degree. In the 
experimental spectrum shown in Fig.\ \ref{fig:cross-sect} (c) this maximum is 
suppressed by the low efficiency for events with small decay energy $E_T$ and 
emission angle $\theta_{\text{Be}}$ approaching zero degree. This is caused by 
the central hole of the second telescope (see Fig.\ \ref{fig:setup}) which, 
however, affects only the low-energy part of the $^{6}$Be spectrum obtained at 
forward angles and does not significantly distort the rest of the spectrum.

The $2^+$ excited state of $^{6}$Be is mainly populated by $\Delta L = 2$ 
momentum transfer. As can be seen from Fig.\ \ref{fig:cross-sect} (c), the 
maximum in the angular distribution obtained for this state is located at 
$\theta_{\text{Be}} = 50^{\circ}-70^{\circ}$, whereas the maximum of the broad 
hump above the $2^+$ state occurs at about $\theta_{\text{Be}}=35^{\circ}$, 
hence at a much smaller angle than that for the maximum of the $2^+$ state. This 
observation suggests that the entire hump structure is related to $\Delta L = 1$ 
transfer populating negative parity states with $J^{-}=\{0^-,1^-,2^-\}$.
To extract more detailed information on this phenomenon the rich data set 
obtained in the experiment was analyzed by taking into account distortions 
caused by the detector array. This task can be performed by using a Monte Carlo 
simulations, provided a proper theoretical model is applied (see Sec.\ 
\ref{sec:theor}).

\section{Theoretical model}
\label{sec:theor}


For the theoretical interpretation of the measured spectra, we extend the 
approach of Ref.\ \cite{Grigorenko:2009c} and combine the three-body dynamics 
for the decay of the $^{6}$Be states with a simple treatment of the reaction 
mechanism. The properties of the $^{6}$Be continuum is approximated by a 
three-body $\alpha$+$N$+$N$ wave function (WF) with outgoing  asymptotics
\begin{equation}
(\hat{H_3}-E_T)\Psi_3^{JM(+)} = \hat{\mathcal{O}}\Psi_{\text{gs}}^{J'M'}\,.
\label{eq:shred}
\end{equation}
The SDM $J^{\pi}=1^-$ continuum in $^{6}$He is assumed to be populated in 
electromagnetic transitions by
\begin{equation}
\Psi_{\text{gs}}\rightarrow \Psi_{^6\text{He}}\,, \quad \hat{\mathcal{O}} \sim
\sum \nolimits _i Z_i r_i \, Y_{1m}(\hat{\mathbf{r}}_i) \,,
\label{eq:d-oper}
\end{equation}
where the index $i$ indicates the constituents which are the $\alpha$-core and 
the two valence nucleons. For the population of low-lying states in $^{6}$He and 
$^{6}$Be in the charge-exchange reaction with $^{6}$Li, the general form of the 
transition operator is given by
\begin{equation}
\Psi_{\text{gs}}\rightarrow \Psi_{^6\text{Li}}\,, \quad \hat{\mathcal{O}} \sim
\sum \nolimits _i f_l(q,r_i) \left[\alpha  + \beta
\sigma_{\mu}^{(i)} \right] \tau_{\pm}^{(i)} Y_{lm}(\hat{\mathbf{r}}_i) \,,
\label{eq:c-e-oper}
\end{equation}
where the index $i$ indicates the two valence nucleons. The operators $\tau_{-}$ 
and $\tau_{+}$ describe the population of the $^{6}$Be and $^{6}$He spectra, 
respectively. Using the effective spin-spin charge-exchange interaction between 
the projectile and target nucleons
\[
V(r) = V_0 (\mathbf{\sigma}^{(1)} \cdot \mathbf{\sigma}^{(2)})
(\mathbf{\tau}^{(1)} \cdot \mathbf{\tau}^{(2)}) \exp[-(r/r_0)^2] \,,
\]
$f_l(q,r_i)$ can be obtained in the plane-wave impulse approximation (PWIA) to 
be
\[
f_l(q,r_i) = V_0 r_0^3 \sqrt{2} \, \pi^2 \exp[-(qr_0/2)^2] \, j_l(qr_i).
\]
Such a simple choice allows one to reproduce qualitatively well the angular 
distributions obtained for $^{6}$Be.

\begin{figure}
\centerline{
\includegraphics[width=0.43\textwidth]{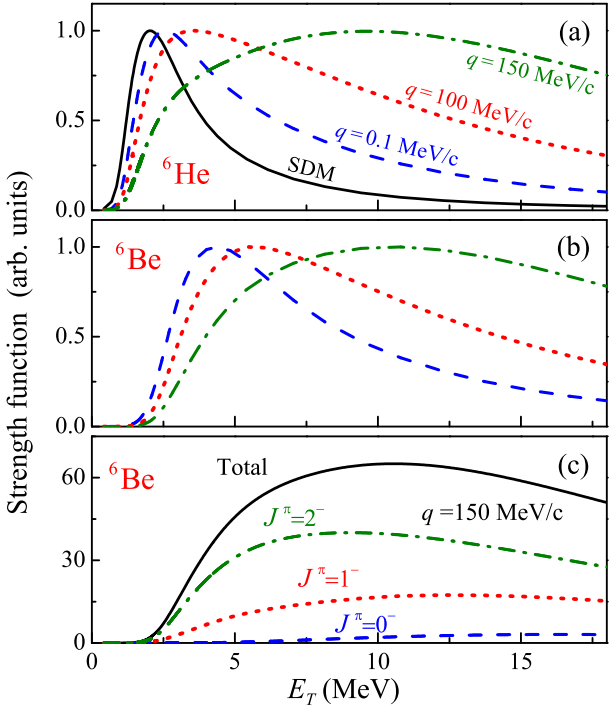}
}
\caption{Theoretical results for negative parity states. (a) The SDM strength
function $dB_{E1}/dE_T$ (solid line) and the IVSDM  spectra ($d\sigma/dE_T$
value) for different transferred momentum $q$ values in $^{6}$He. (b) The IVSDM
($d\sigma/dE_T$ value) spectra in $^{6}$Be. The meanings of the curves is the
same as
in panel (a).  All curves are normalized to the unity maximum value. Panel (c)
shows the contributions of different $J$ values in one selected ($q=150$ MeV/c)
IVSDM spectrum of $^{6}$Be.}
\label{fig:soft-dipole}
\end{figure}

The three-body cluster model \cite{Danilin:1991}, developed for the $A = 6$ 
isobaric chain, was applied to calculate the g.s.\ wave-function 
$\Psi_{\text{gs}}$. The method, which was successfully applied in Ref.\ 
\cite{Grigorenko:2009} to the studies of the $0^+$ state, was used for the 
resonant $0^+$ and $2^{+}$ states. The calculated cross section profiles are 
shown in Fig.\ \ref{fig:cross-sect} (b). The cross-section ratios are obtained 
by fitting procedures, as will be discussed in Section \ref{sec:angular}.

The continuum wave-function of negative parity states were obtained by using the 
Green's function method \cite{Grigorenko:2006}
\begin{equation}
\Psi_3^{JM(+)} = \hat{G}^{(+)}_{3E_T}\hat{\mathcal{O}}\Psi_{\text{gs}}^{J'M'}\,,
\label{eq:wf-plus}
\end{equation}
where the exact Green's function of a simplified three-body Hamiltonian is used 
which is available in compact analytical form:
\begin{equation}
\hat{G}^{(+)}_{3E}(\mathbf{XY},\mathbf{X}'\mathbf{Y}') = \frac{E}{2\pi i}
\int^{1}_0 \!\!\! d \varepsilon \,
\hat{G}^{(+)}_{\varepsilon E}(\mathbf{X},\mathbf{X}')
\hat{G}^{(+)}_{(1-\varepsilon)E}(\mathbf{Y},\mathbf{Y}').
\label{eq:gf3}
\end{equation}
The vectors $\mathbf{X}$ and $\mathbf{Y}$ are the Jacobi variables for the ``Y'' 
system: $\mathbf{X}$ connects the core with one of the valence nucleons, while 
$\mathbf{Y}$ connects the second nucleon with the center of mass of the remnant. 
The operator $\hat{G}^{(+)}_{E}$ is an ordinary two-body Green's function. This 
method takes into account exactly one final-state interaction out of the three 
ones being present. This was shown to be a good approximation in the case of SDM 
\cite{Grigorenko:2006,Golovkov:2009}, where the dynamics is defined 
predominantly by one resonant subsystem (for A=6 nuclei this is evidently the 
$p_{3/2}$  resonance in the $\alpha$-$N$ subsystem). An analogous approximation 
has been used in Ref.\ \cite{Esbensen:1992} for the SDM studies in $^{11}$Li.

Figures \ref{fig:soft-dipole} (a,b) show the comparison of the strength 
functions for the SDM in $^{6}$He, see Eq.\ (\ref{eq:d-oper}), and the IVSDM in 
$^{6}$He and $^{6}$Be built on $^{6}$Li, see Eq.\ (\ref{eq:c-e-oper}). As can be 
seen from Fig.\ \ref{fig:soft-dipole}, the peak of the IVSDM of $^{6}$He for 
small $q$ values occurs close to that of the SDM but is much broader. This is a 
consequence of the more compact g.s.\ wave-function of $^{6}$Li as compared to 
$^{6}$He. However, the steep rise of the spectra occurs in the case of the 
$^{6}$He IVSDM at energy $1 - 2$ MeV comparable with the SDM case. As can be 
seen from Fig.\ \ref{fig:soft-dipole} (b), the low-energy slope of the IVSDM in 
isobaric mirror partner of $^{6}$He (namely $^{6}$Be) appear to be at higher 
energies of $2 - 4$ MeV. The theoretical IVSDM curves for $q = 150$ and $q = 
100$ MeV/c, are displayed in Fig.\ \ref{fig:soft-dipole} (b) for $^{6}$Be. These 
$q$ values correspond to c.m.\ angles of about $\sim 45 ^{\circ}$ and $\sim 28 
^{\circ}$ where the experimental IVSDM peak is located, see Fig.\ 
\ref{fig:cross-sect} (c). Thus $q \sim 100 - 150$ MeV/c is a typical momentum 
transfer for the IVSDM peak. Calculations for $q = 0.1$ MeV/c can be considered 
to represent the ``long-wave limit''. In this case the transition operator 
approximately behaves like $\sim r_i$ providing the dynamical analogue of SDM 
calculations. The ``long-wave'' and ``short-wave'' responses have quite 
different shapes. However, the phenomenon has stable features with respect to 
assumptions concerning the source term (reaction mechanism). For example, the 
strength functions rise practically at the same energy. One can also see that 
the IVSDM contributes already at the energy of the $2^+$ state, i.e.\ around 3 
MeV.

Figure \ref{fig:soft-dipole} (c) shows the population of different $J^-$ states 
in our model. The populations of $2^-$ and $1^-$  dominate whereas the 
contribution of $0^-$ is weak. On the one hand, the presence of states with 
negative parity but with different $J$ values makes the picture more complex 
than in the case of the SDM in $^{6}$He, where only the $J^{\pi}=1^-$ state is 
believed to be populated in the Coulomb excitation. On the other hand, the IVSDM 
provides access to a wider range of dynamical phenomena which can represent an 
important feature of such studies.


\section{Data analysis: correlations in $^{6}$Be }
\label{sec:ang-cor}


\begin{figure*}[t]
\centerline{
\includegraphics[width=0.272\textwidth]{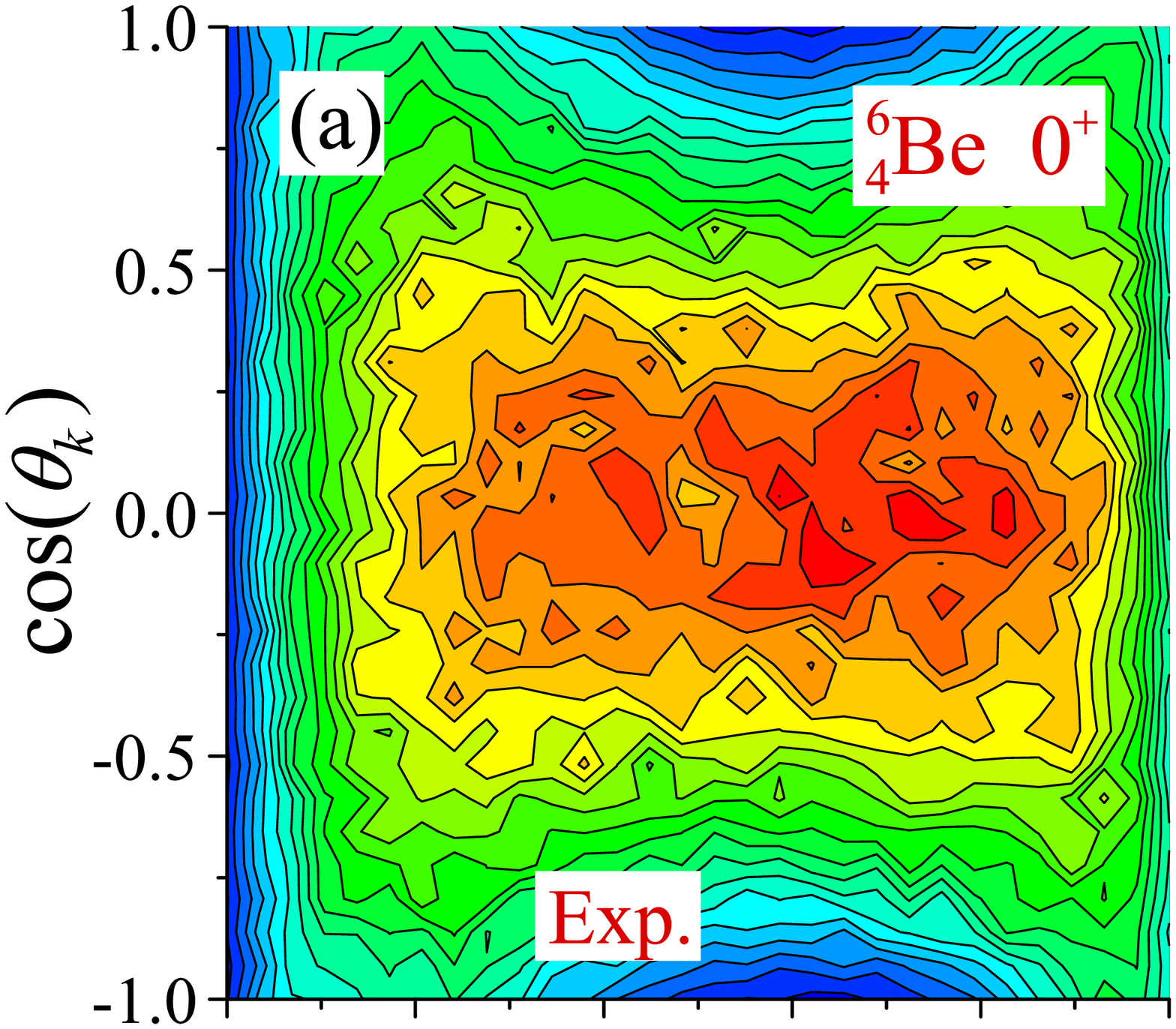}    
\includegraphics[width=0.23\textwidth]{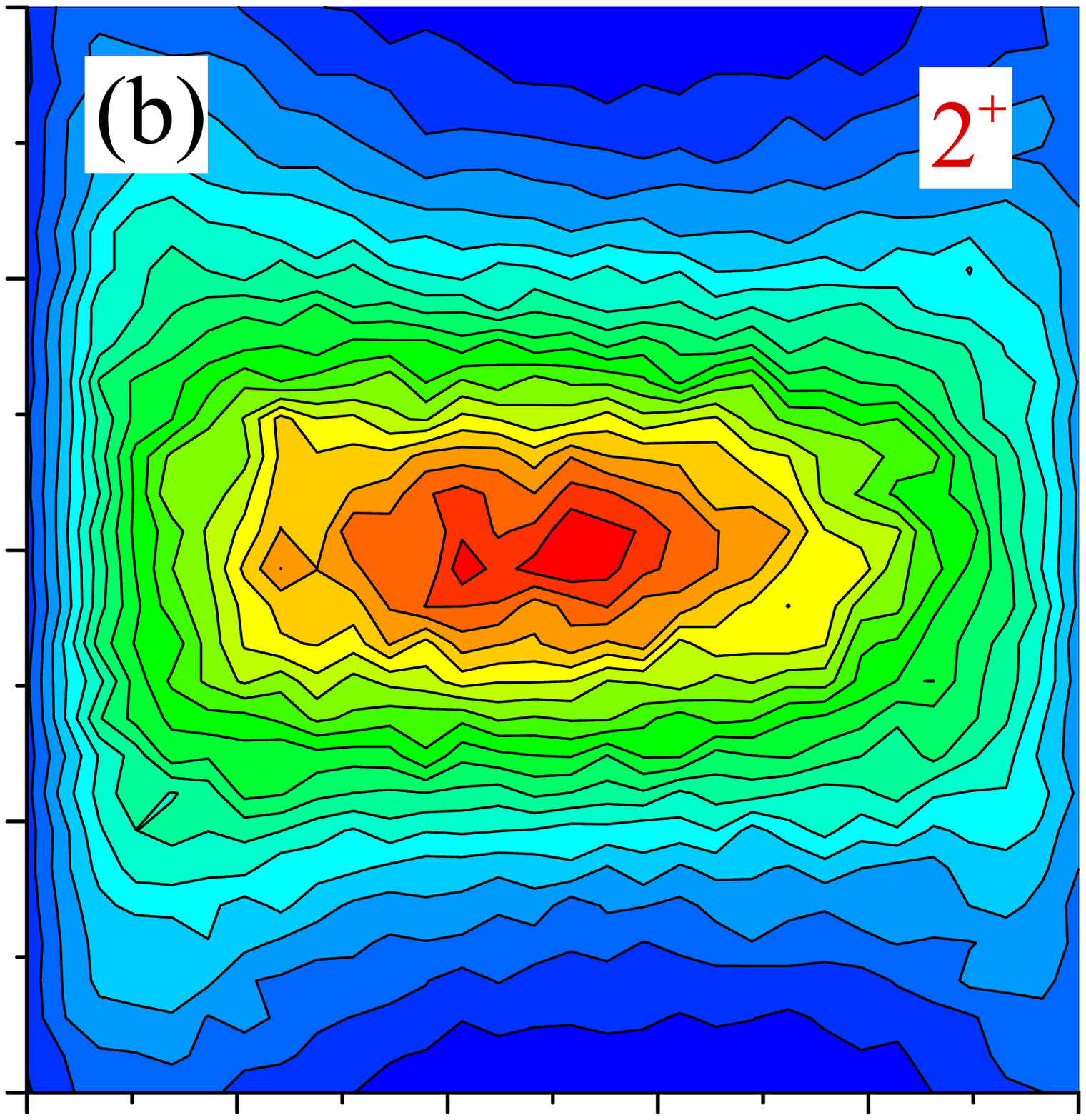}    
\includegraphics[width=0.23\textwidth]{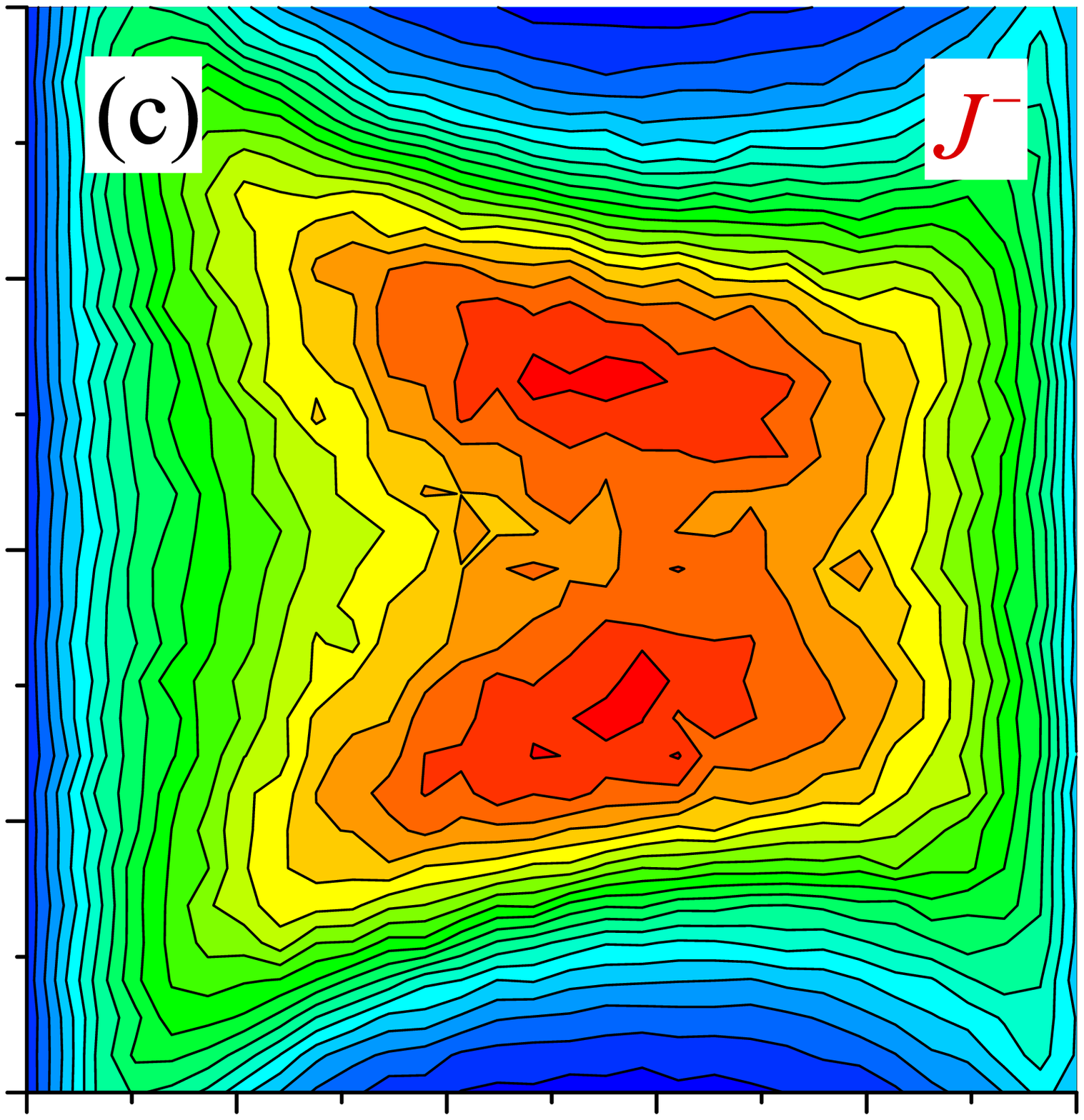}   
\includegraphics[width=0.23\textwidth]{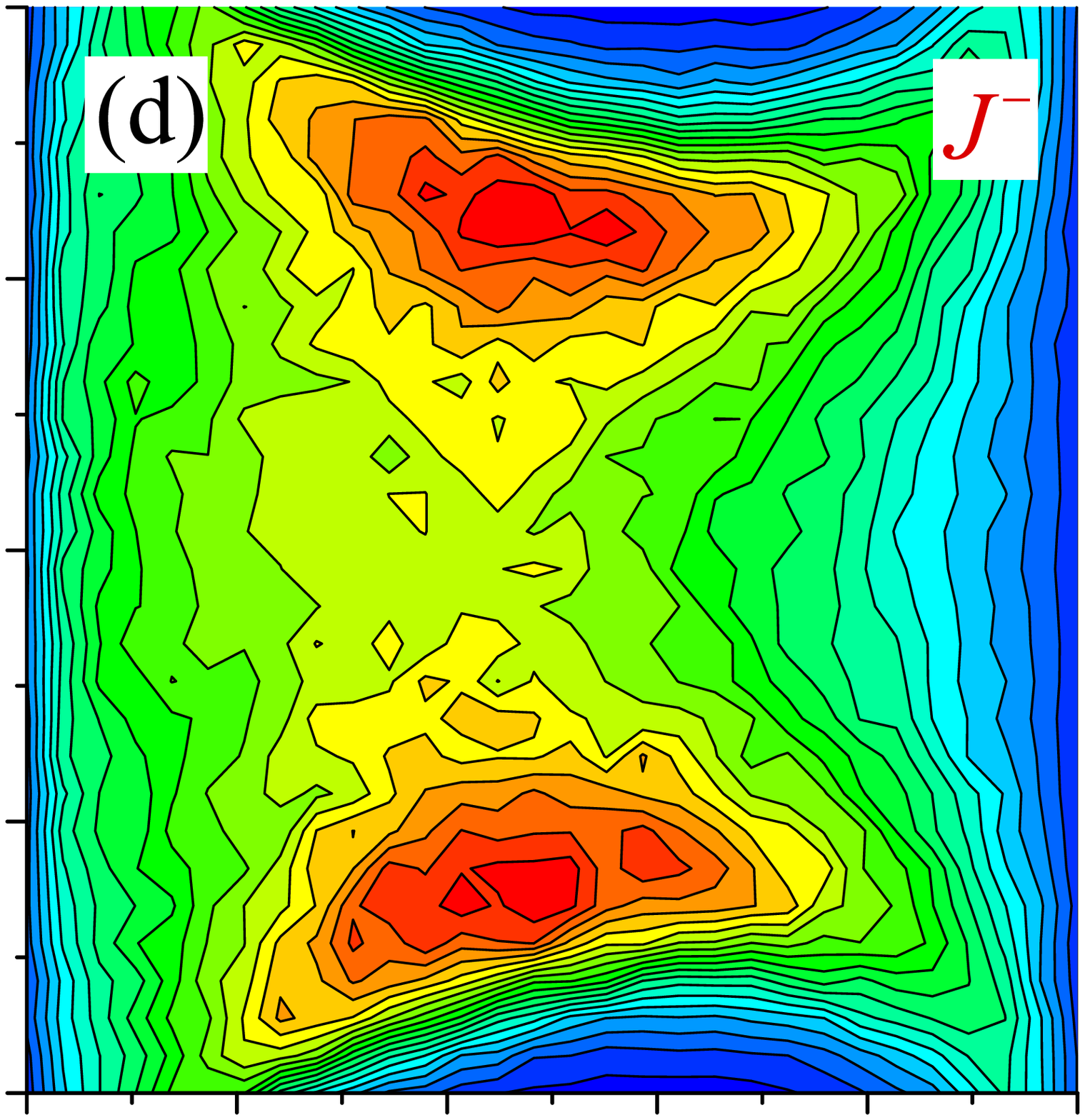}   
}
\centerline{
\includegraphics[width=0.274\textwidth]{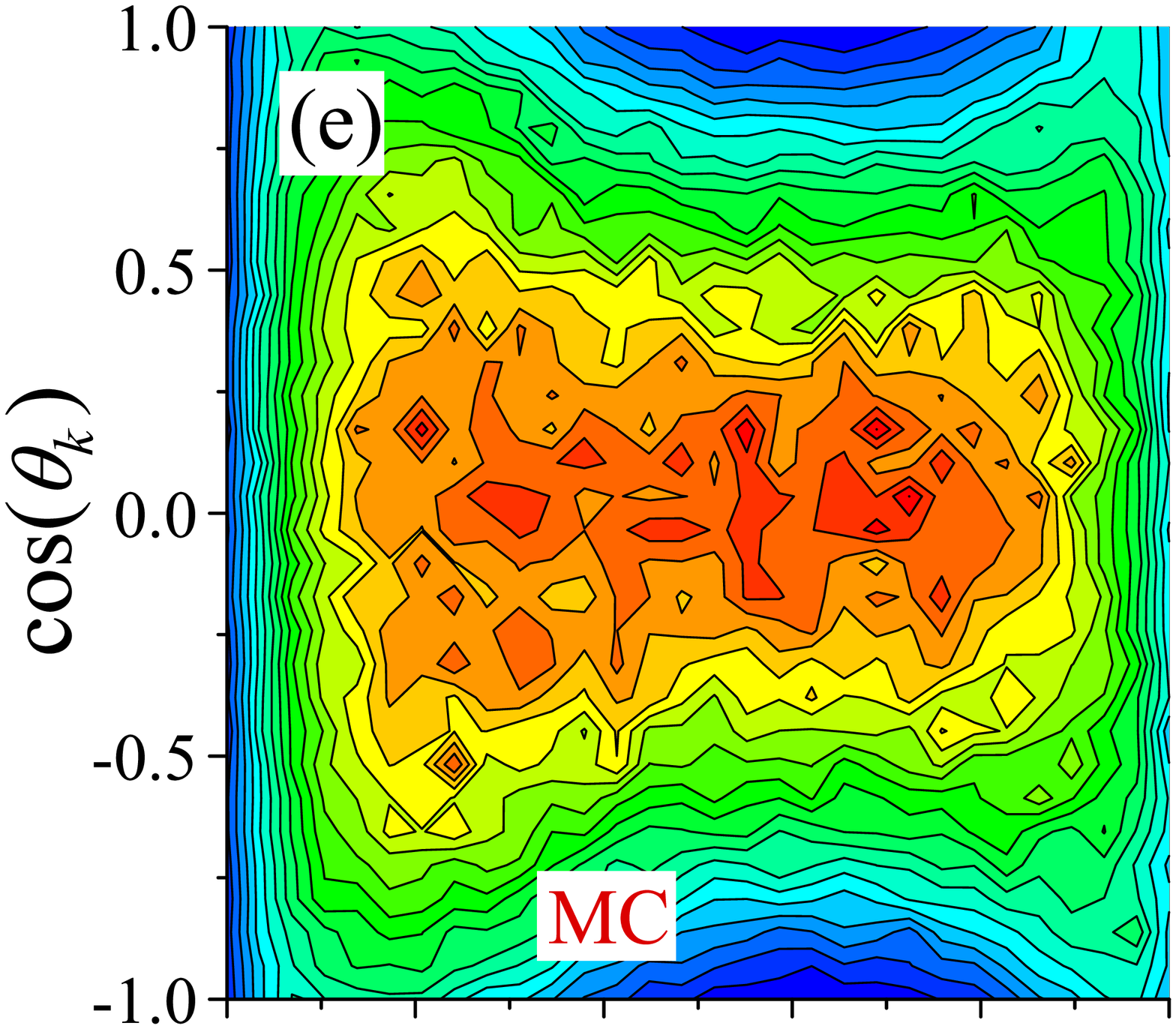}      
\includegraphics[width=0.23\textwidth]{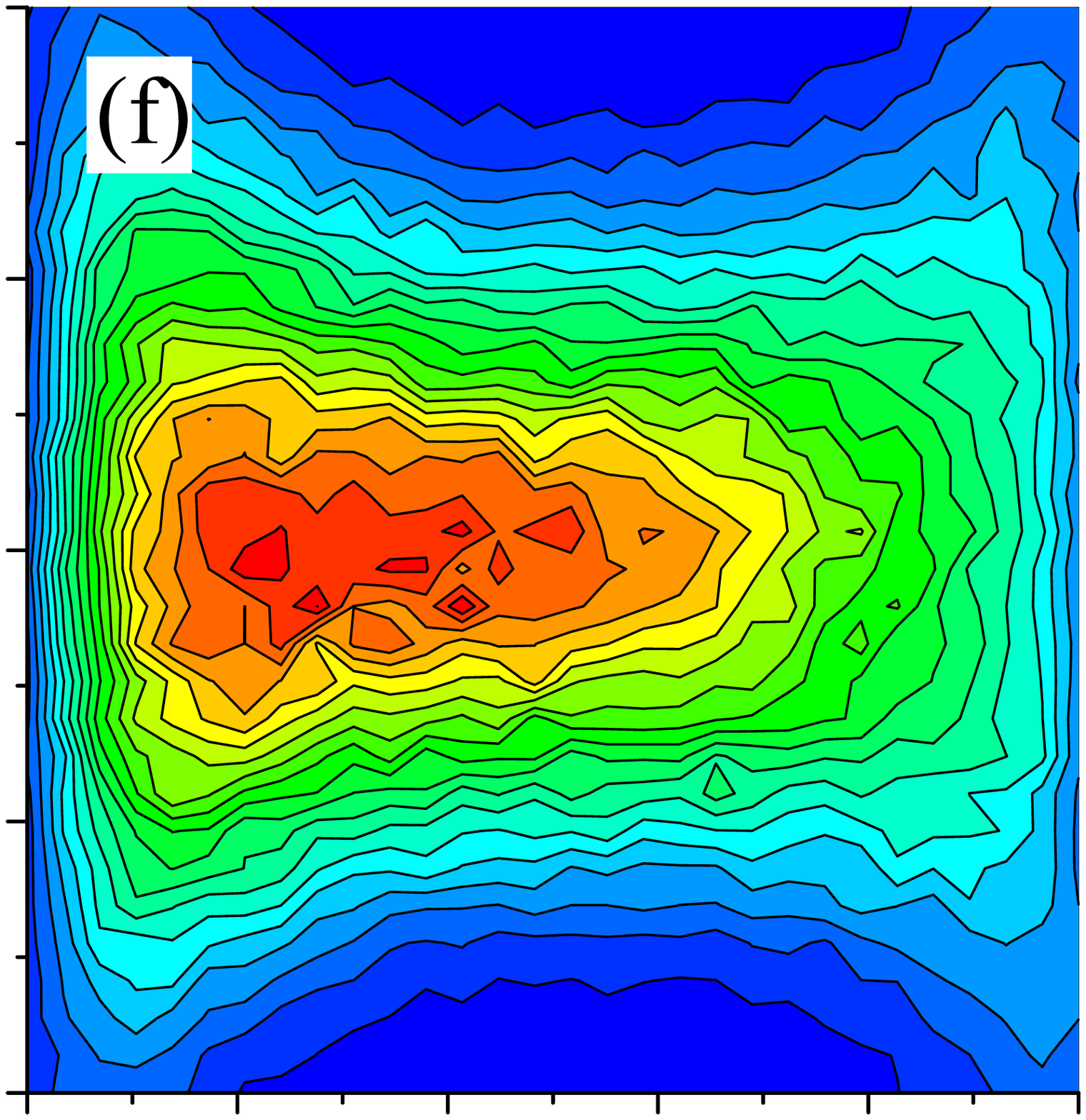}      
\includegraphics[width=0.23\textwidth]{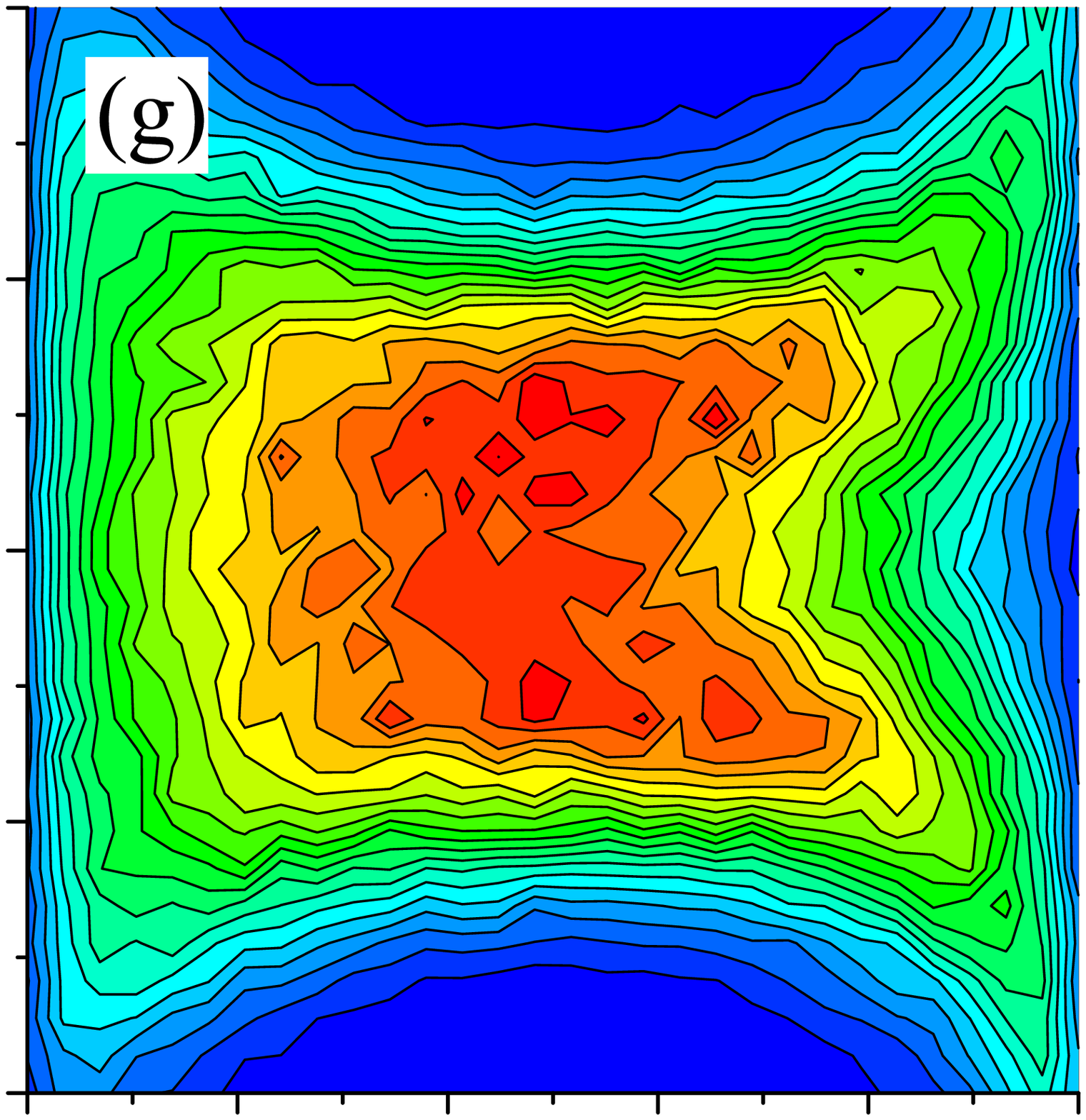}     
\includegraphics[width=0.23\textwidth]{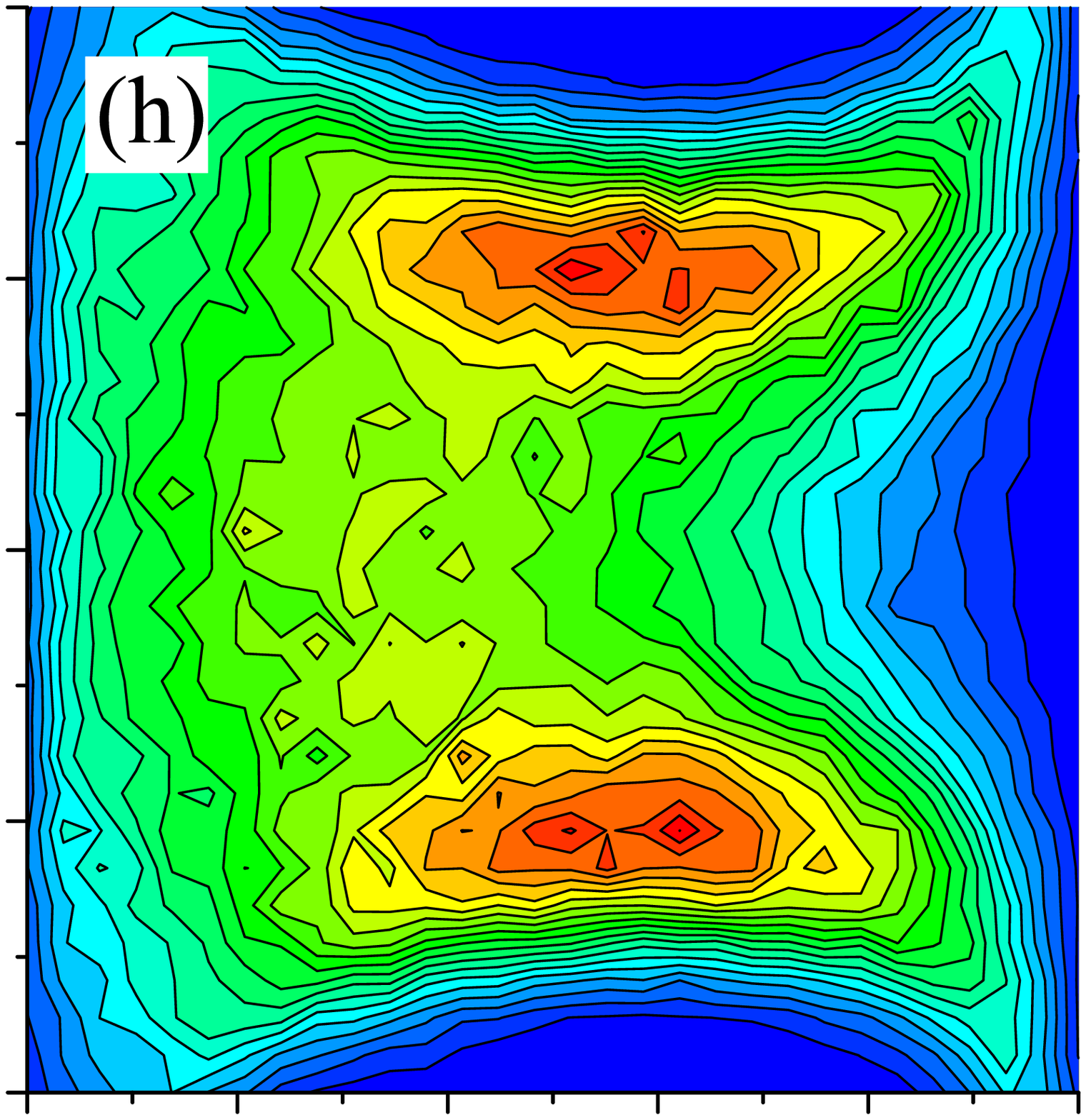}     
}
\centerline{
\includegraphics[width=0.274\textwidth]{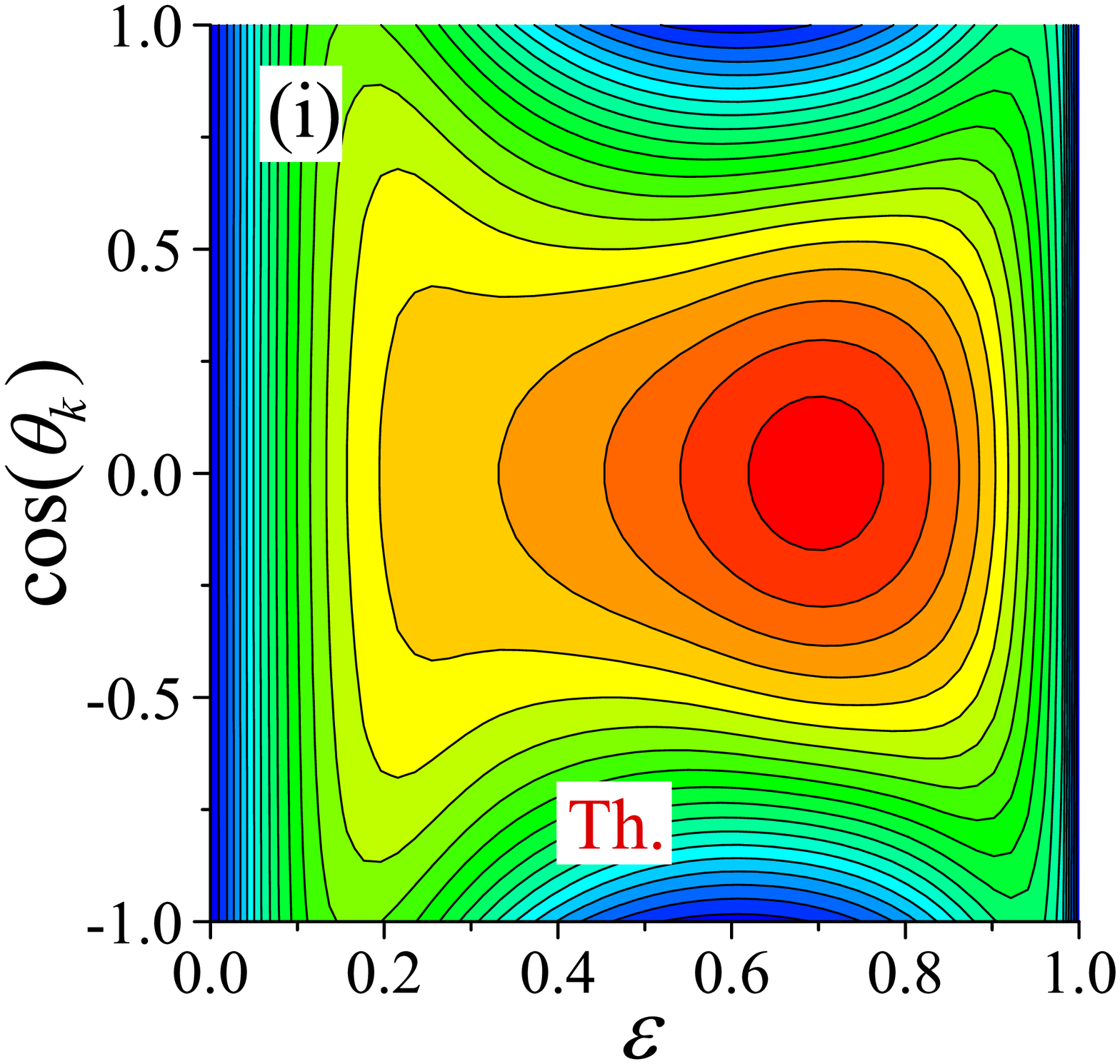}      
\includegraphics[width=0.23\textwidth]{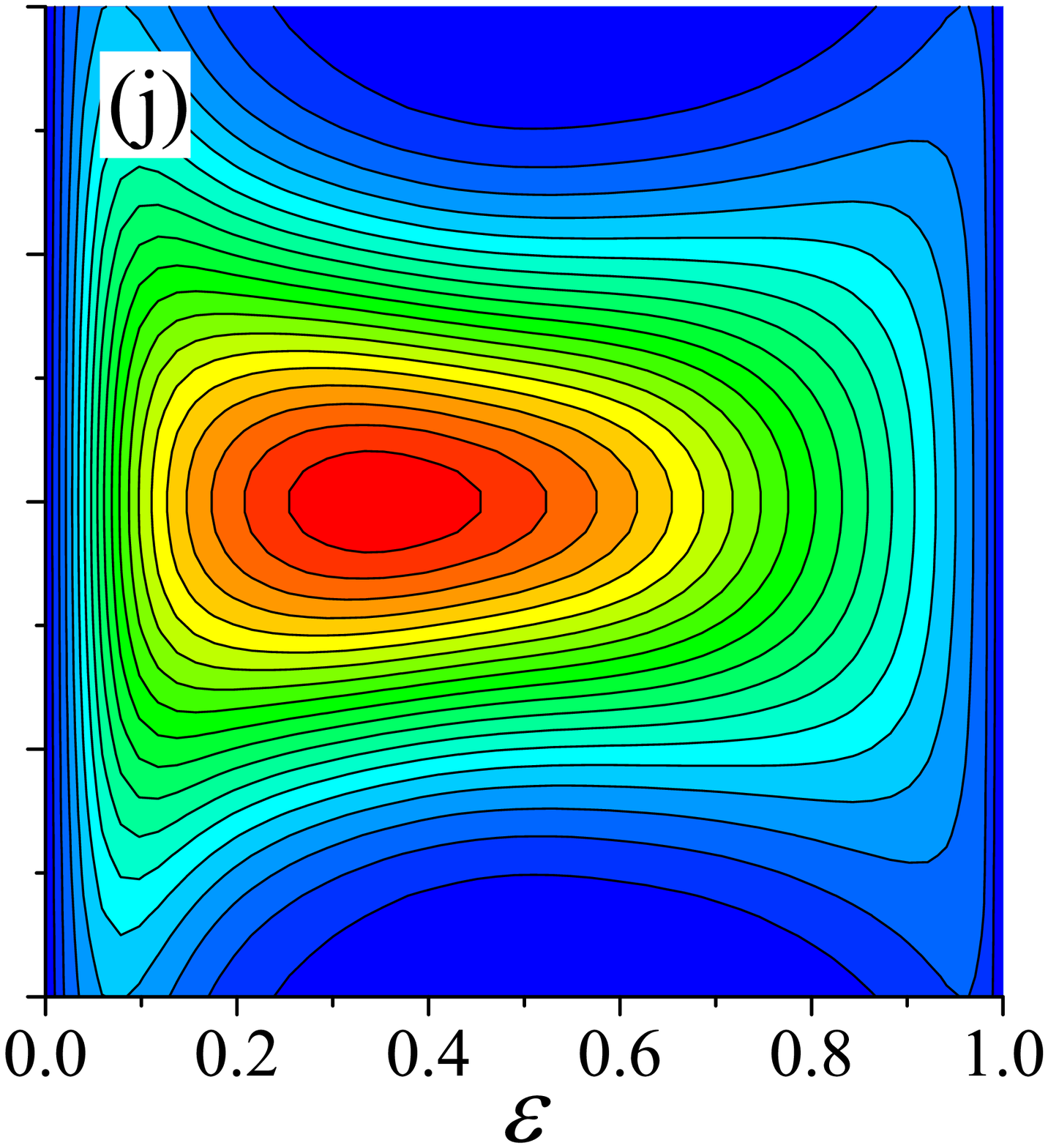}      
\includegraphics[width=0.23\textwidth]{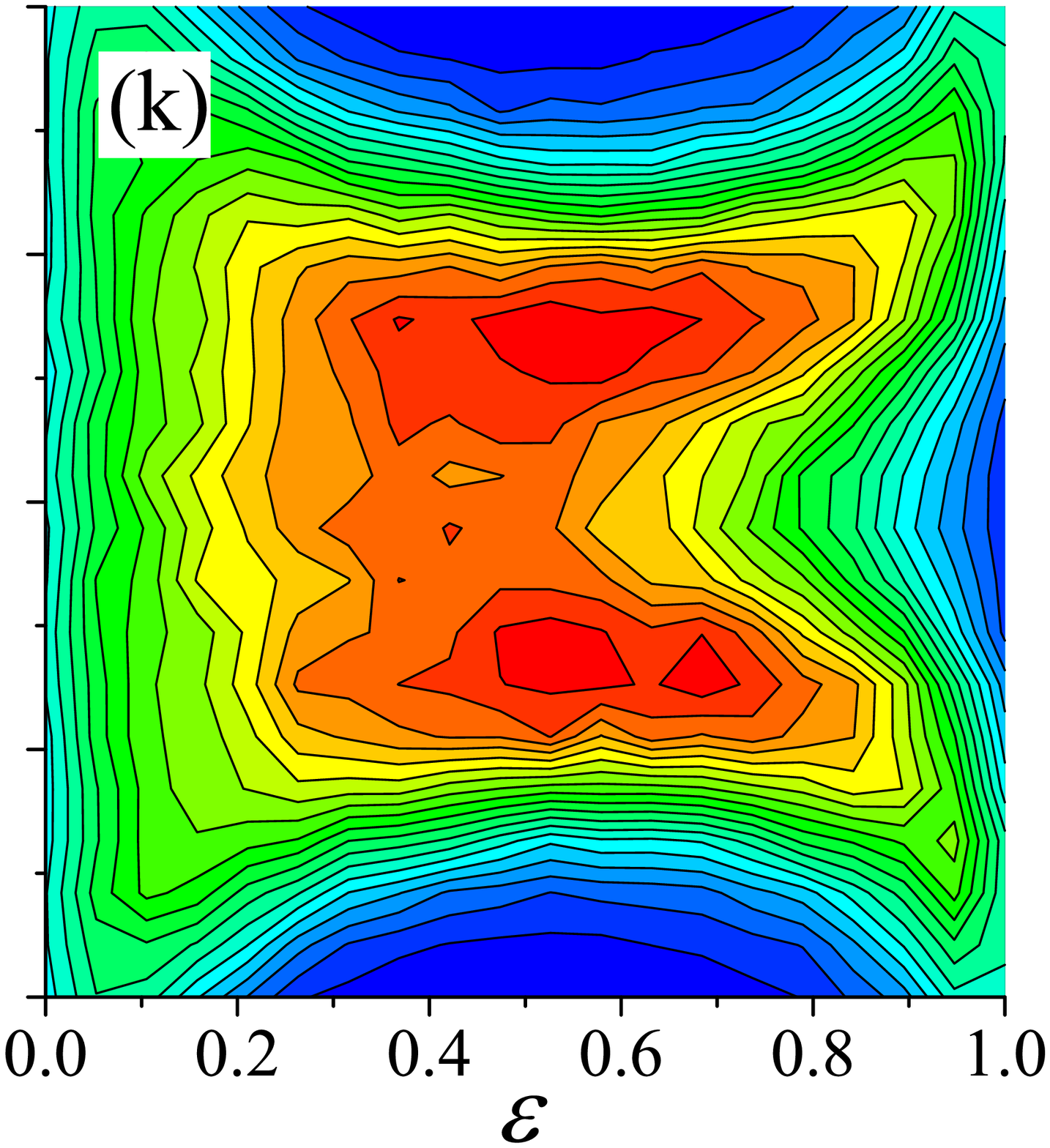}     
\includegraphics[width=0.23\textwidth]{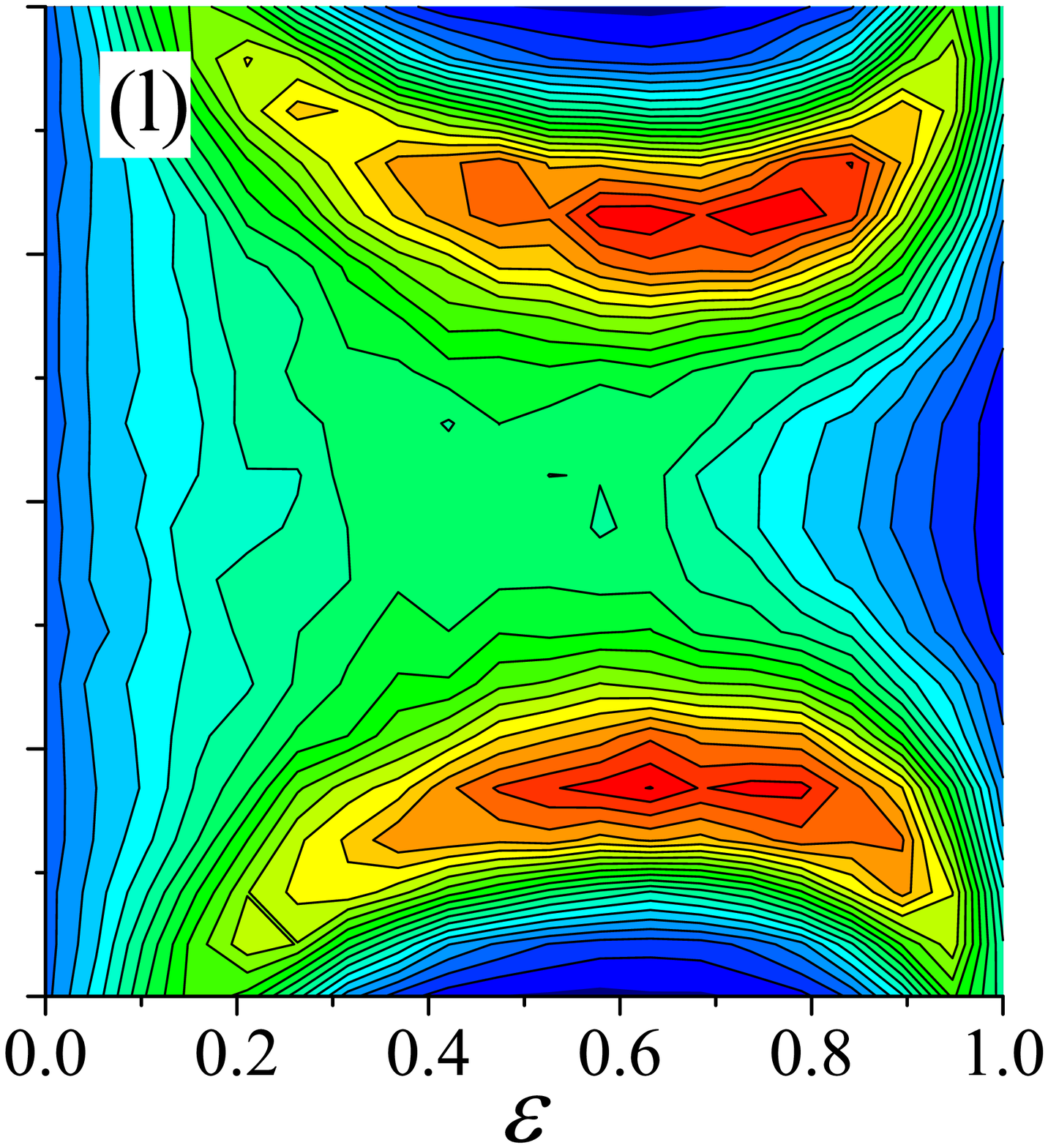}     
}
\caption{Internal energy-angle correlations for different states of $^{6}$Be are 
presented in Jacobi ``T'' system. Compared are the experimental (upper row), 
Monte Carlo (middle row) and theoretical (lower row) correlation patterns. 
Results are presented for the resonant $0^+$ (a,e,i) and $2^+$ (b,f,j) states. 
The $J^-$ configurations of the IVSDM are given for two energy ranges, namely 5 
to 7 MeV (c,g,k) and 9 to 11 MeV (d,h,l). The kinematical variables are defined 
in Eq.\ (\ref{eq:corel-param}).}
\label{fig:comp-mom-dis}
\end{figure*}

Without taking the initial orientation into account, the energy-angle 
correlations of the fragments in the $^{6}$Be c.m.\ (so-called ``internal'' 
three-body correlations) can be described in terms of the energy distribution 
parameter $\varepsilon $ and the angle $\theta_{k}$ between the Jacobi momenta 
$\mathbf{k}_{x}$, $\mathbf{k}_{y}$,
\begin{eqnarray}
\varepsilon = E_x/E_T \quad ,\quad \cos(\theta_k)=(\mathbf{k}_{x} \cdot
\mathbf{k}_{y}) /(k_x\,k_y) \, , \nonumber \\
E_T =E_x+E_y=k^2_x/2M_x + k^2_y/2M_y \, , \nonumber \\
{\bf k}_x  =  \frac{A_2 {\bf k}_1-A_1 {\bf k}_2 }{A_1+A_2} \, ,  \,\;
{\bf k}_y  =  \frac{A_3 ({\bf k}_1+{\bf k}_2)-(A_1+A_2) {\bf k}_3}
{A_1+A_2+A_3}\,, \;
\label{eq:corel-param}
\end{eqnarray}
where $M_x$ and $M_y$ are the reduced masses of the $M_x$ and $M_y$  subsystems 
(see, e.g., Ref.\ \cite{Grigorenko:2009c} for details).

To compare the model calculations with the experimental results we made two 
additional assumptions, namely we neglected the spin alignment of $^{6}$Be and 
interference effects. The upper row of Fig. 5 shows the energy-angle 
correlations obtained for $^{6}$Be decay products. The correlation plots are 
built in the ``T'' system from the experimental data taken in four different 
ranges of $E_T$. The first column of Fig.\ \ref{fig:comp-mom-dis} corresponds to 
the g.s.\ population in $^{6}$Be, the second column shows the correlations 
obtained in the energy range where the first excited $2^+$ state has maximal 
cross section, and  the  last  two columns display similar correlations revealed 
for $E_T$ ranging from 5 to 7 MeV and 9 to 11 MeV, respectively, where the 
population of negative-parity states dominates. The lowest row in Fig.\
\ref{fig:comp-mom-dis} shows theoretical calculations for the energy-angle 
correlations obtained in the same energy regions. The middle row shows the 
results of Monte Carlo simulations which were performed in order to correct the 
theoretical calculations for the instrumental bias. This row can be compared 
with the upper one where the experimental results are displayed.

Correlations similar to those presented for the $^{6}$Be g.s.\ in the first 
column of Fig.\ \ref{fig:comp-mom-dis}  were studied in detail in Refs.\
\cite{Grigorenko:2009,Grigorenko:2009c}, where the experimental data were well 
reproduced by theoretical calculations. Our experimental data for the $0^+$ 
state also yield full agreement with the model calculations and thus with the 
data of Ref.\ \cite{Grigorenko:2009c}. In the present work, high quality 
energy-angle correlation data were obtained for the $2^+$ state of $^{6}$Be for 
the first time. The data deduced for the $2^+$ state are qualitatively different 
from those measured for the $0^+$ state: the profile of the $\cos(\theta_k)$ 
distribution is much more narrow for the $2^+$ state, and the maximum of the 
correlation distribution is shifted towards smaller relative proton energy. The 
former observation is consistent with the predictions made for the 
``democratic'' decays in Ref.\ \cite{Pfutzner:2012}. The sensitivity of the 
three-body correlations to the underlying nuclear structure indicates that they 
may represent a tool for spin-parity identification. The properties of the $2^+$ 
state will be discussed to more detail in a forthcoming publication.

The correlations of the broad hump at $E_T>4$ MeV are qualitatively different 
from the correlations of the $0^+$ and $2^+$ state. Correlations found for 
(tentatively assigned) $J^-$ states show a smooth dependence on the $^{6}$Be 
energy but almost none on the c.m.\ angle $\theta_{\text{Be}}$. The pronounced 
crescent-like ridges exhibited in columns 3 and 4 of Fig.\ 
\ref{fig:comp-mom-dis} are connected to the $p_{3/2}$ and $p_{1/2}$ resonant 
final state interactions in the core-$p$ subsystem. The evolution of these 
ridges with increasing $E_T$ is well described under the assumption of the 
IVSDM. Comparing the first and second rows in Fig.\ \ref{fig:comp-mom-dis} one 
can see some differences in details which, most probably, can be explained by 
interference effects. We leave this problem for a further analysis but emphasize 
the overall agreement between model calculations and experimental correlations.


\section{Data analysis: angular distributions and partial cross sections}
\label{sec:angular}


In order to analyse the angular distributions and the partial cross-sections, 
the data displayed in Fig.\ \ref{fig:cross-sect} (c) were divided into narrow 
angular intervals and projected on the energy axis. By using the theoretically 
predicted energy profiles of the partial cross sections for different $J^{\pi}$ 
values, the experimental data were fitted for each angular interval, yielding 
the angular distributions for individual $J^{\pi}$ contributions. The $q$ 
dependence of the partial cross sections was neglected, and the cross sections 
were calculated by using the $q$ values defined for each  $J^{\pi}$ by the 
maximum in the angular distribution. Nevertheless, performing Monte Carlo 
simulations based on the results obtained by the theoretical model, we are able 
to well reproduce the spectrum displayed in Fig.\ \ref{fig:cross-sect} (c) in 
the whole energy and angular ranges. The shaded histogram shown in Fig.\ 
\ref{fig:cross-sect} (a) was accumulated by summing the local projection. The 
reconstructed $^{6}$Be spectrum, corrected for the instrumental efficiency and 
integrated over the whole angular range, is shown in Fig.\ \ref{fig:cross-sect} 
(b). The energy profiles predicted by theory for the $0^+$, $2^+$ and $J^-$ 
states were found to be consistent with the experiment, and the ratios of 
reaction yields were determined for these states.

\begin{figure}
\centerline{
\includegraphics[width=0.303\textwidth]{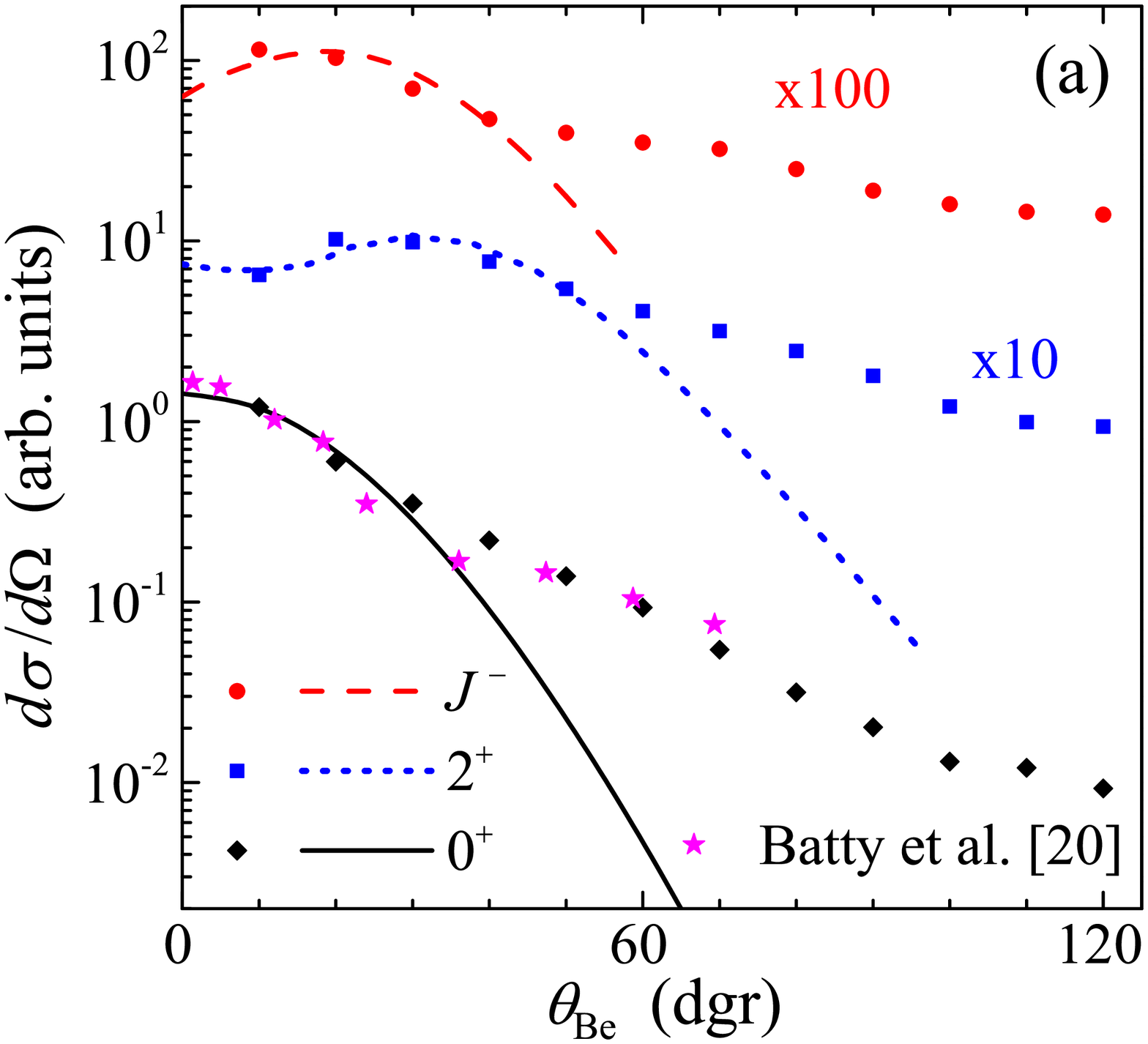}
\includegraphics[width=0.188\textwidth]{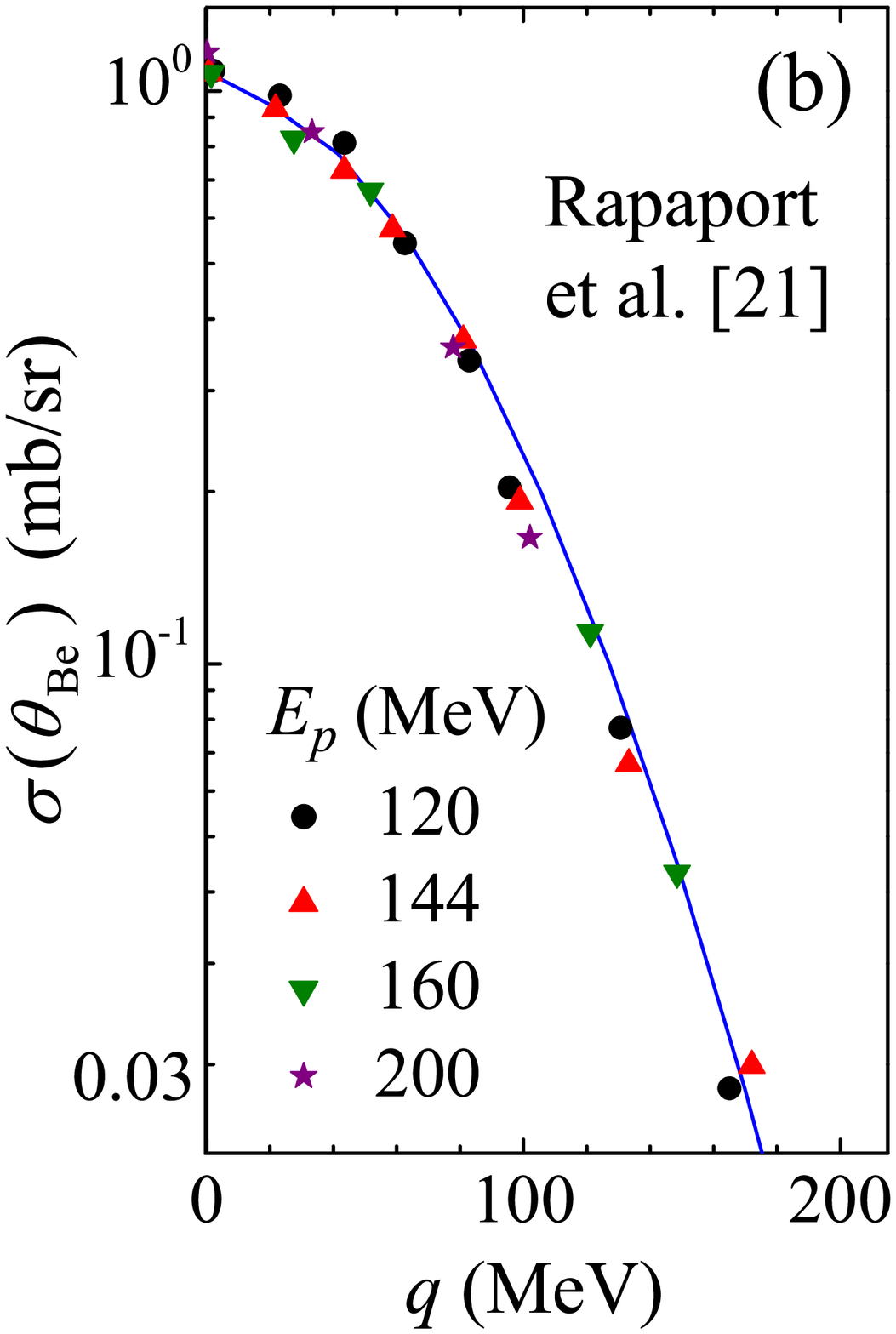}
}
\caption{(a) Angular distributions for states with different $J^{\pi}$. Symbols 
show the experimental data, and curves of the same color provide PWIA results. 
The $0^+$ and $2^+$ data are those obtained on resonance while the IVSDM $J^-$ 
distribution is integrated over energy. The data of Ref.\ \cite{Batty:1968} for 
the $^{6}$Be g.s.\ are shown for comparison (pink stars). (b) Dependence of the 
g.s.\ cross-section on the transferred momentum calculated in PWIA (solid curve) 
and compared to the data obtained in Ref.\ \cite{Rapaport:1990}  at different 
proton energies $E_p$ (symbols).}
\label{fig:ang-distr}
\end{figure}

The resulting angular distributions for the $^{6}$Be states, as populated in the 
$^1$H($^6$Li,$^6$Be)$n$ reaction, are displayed in Fig.\ \ref{fig:ang-distr} 
(a). The angular distribution for the $^{6}$Be g.s.\ is in a good agreement with 
that obtained by using a $^{6}$Li($p$,$n$) reaction  \cite{Batty:1968} at a 
proton beam energy of 30.2 MeV, which yields a similar c.m.\ energy. The latter 
data were evaluated by using the missing-mass method, without applying any 
further data handling such as a complex efficiency corrections. The fact that 
the ($p$,$n$) results agree well with those obtained in our work is a strong 
confirmation of our data treatment. To calibrate the PWIA calculations we used 
the $^{6}$Li($p$,$n$) data from Ref.\ \cite{Rapaport:1990}. The distributions of 
 transferred momenta, measured for several incident proton energies, are well 
reproduced by using an $r_0$ parameter of 0.33 fm [see the solid curve in Fig.\ 
\ref{fig:ang-distr} (b)]. Calculations with this $r_0$ value also reproduce the 
peak positions of different components of the $^6$Be spectrum, obtained in our 
experiment [see Fig.\ \ref{fig:ang-distr} (a)].

The PWIA calculations indicate that the population of the $0^+$ and $2^+$ is 
associated mainly with $\Delta L=0$ and $\Delta L=2$ transitions, respectively. 
The $2^+$ state also has an important $\Delta L=0$  contribution which is 
responsible for the rise of the cross section near the forward direction. The 
broad hump at $E_T>4$ MeV is apparently associated with $\Delta L=1$. 
Qualitatively, the theoretical angular distributions are very similar to the 
experimental ones. However, they do not reproduce the tails of the measured 
distributions which extend to large c.m.\ angles. This deficiency is evidently 
connected to our simplified treatment of the reaction mechanism (PWIA), see 
discussion in Refs.\ \cite{Batty:1968,Petrovich:1993}. It is not expected that 
this deficiency affects the conclusions drawn from the correlation data. The 
evident correspondence of the measured angular distributions with $\Delta L$ and 
the reasonable description of the experimental data by a simple PWIA calculation 
support, firstly, the dominance of the direct reaction mechanism in our 
experiment and, secondly, the overall consistency of the our data 
interpretation.

\section{Major qualitative features of the IVSDM}


To properly describe the two-body final state interactions within the framework 
of the three-body continuum remains a challenge for theoreticians. In view of 
the systematic disagreement between different calculations 
\cite{Cobis:1997,Danilin:1998,Myo:2001}, it is evident that there is a serious 
problem in interpreting the soft E1 mode of two-neutron halo nuclei. In order to 
avoid this complexity we used a model for describing the soft responses which is 
simplified but presumably contains the major features of the relevant phenomena.

Before discussing the IVSDM from a theoretical point of view, we emphasize 
several formal differences between the SDM and the IVSDM:

\noindent(a) In $N = Z$ nuclei the E1 dipole transitions to states with the same 
isospin are strongly suppressed (e.g., by 3 orders of the magnitude) due to the 
small effective dipole charges of the nucleons. Therefore, the continuum 
population by $\Delta T=0$ transitions (like the SDM in $^{6}$He) is not 
possible in $^{6}$Li.

\noindent(b) Electromagnetic isovector dipole transitions in $^{6}$Li (e.g.\ 
Ref.\ \cite{Berman:1975}) are induced by the spin-scalar operator, while in the 
strong isovector transitions the spin-vector part of the transition is expected 
to dominate. These operators correspond in Eq.\ (\ref{eq:c-e-oper}) to the terms 
with coefficients $\alpha$ and $\beta$, respectively, and populate distinctly 
different configurations in the final state.

\noindent(c) In $^{6}$He, the soft E1 transition to continuum is due to the 
interaction with the charged core \emph{only}. In the case discussed here, the 
charge-exchange reaction occurs with the valence nucleons \emph{only}. Therefore 
the IVSDM is expected to be sensitive to different nuclear-structure features 
than the SDM.

We consider the following qualitative features  to be exclusive for our 
interpretation for the observed phenomenon:

\noindent(i) The proposed phenomenon represents a ``major effect''. It can be 
seen in Fig.\ \ref{fig:cross-sect}, that it is responsible for  the major part, 
i.e.\ about $60\%$, of the  charge-exchange cross-section populating the 
$^{6}$Be states up to $E_T = 16$ MeV. It is a challenging task to try to 
interpret the dominant population of the non-resonant (negative parity) 
continuum, in comparison with the resonance continuum states.

\noindent(ii) The energy and angle dependence of the IVSDM cross-section depends 
on details of the reaction mechanism. In particular, in PWIA calculations it 
sizably depends on the transferred momentum $q$. Most likely, a qualitatively 
similar dependence exists in models for other reactions. This $q$ dependence 
means that an attempt to characterize the IVSDM in terms of resonances with 
definite energies and widths does not take the underlying physics properly into 
account. The term ``excitation mode'' characterises this phenomenon correctly as 
it comprises a large observable effect which, however, is sensitive to the 
specific observation conditions.

\noindent(iii) The $^{6}$Be g.s.\ is particle-unstable and thus the strong 
population of the negative parity continuum can not be interpreted as being due 
to the soft excitations based on the weakly-bound and therefore 
radially-extended ground state. We interpret the observed phenomenon as having 
two roots, namely the radially extended g.s.\ of $^{6}$Li (not quite as large in 
$^{6}$He) and the final state interactions in the negative parity continuum of 
$^{6}$Be.

\noindent(iv) The case of the IVSDM gives opportunity to test the concept of 
isobaric symmetry for this class of phenomena. According to the calculations 
shown in Fig.\ \ref{fig:soft-dipole}, the IVSDM is a ``threshold-oriented'' 
phenomenon. This means that the Coulomb effect can be observed as a dependence 
on \emph{threshold energy} $E_T$, corresponding to a shift of the $^{6}$Be 
distributions to higher $E_T$ values in comparison to the $^{6}$He data. 
However, the \emph{excitation energies} $E^*$ (counted from the respective 
ground states) of the peaks of the IVSDM distribution, are about the same for 
$^{6}$He and $^{6}$Be and maybe even somewhat lower for $^{6}$Be, which is quite 
unexpected.


\section{Discussion of experiments related to the IVSDM}
\label{sec:disc}


The standard compilation \cite{Tilley:2002} including also the results obtained 
in charge-exchange reactions of the type $(p,n)$ and $(^3$He$,t)$, populating 
the $^{6}$Be continuum, yields an ambiguous situation concerning the data on the 
$^{6}$Be excitation energies $E_T$ between 1.67 and 23 MeV. Older studies 
\cite[and Refs.\ therein]{Givens:1972,Geesaman:1977,Bochkarev:1992} were 
confined to narrow angular ranges. For certain angular ranges, excitation 
spectra similar to ours have been obtained \cite{Batty:1968,Bochkarev:1992}. In 
all these cases, however, the spin-parity identification was not possible and 
excitations above the $2^+$ state have generally been interpreted as being due 
to the three-body $\alpha$+$p$+$p$ ``phase volume''.

In Ref.\ \cite{Yang:1995}, the $^{6}$Be spectrum was populated by using the 
$(p,n)$ reaction at high proton energy, namely 186 MeV. Above the $2^+$ state of 
$^{6}$Be, four groups of events with important $\Delta L = 1$ components were 
distinguished, i.e.\ at $E^*\sim 5.5$, 10, 15, and 25 MeV. According to the 
interpretation given in Ref.\ \cite{Yang:1995}, the cross section in the range 
$E^*\sim 3-16$ MeV has a significant $\Delta L = 0$  contribution, ranging from 
about $50\%$ at $E^*\sim 3$ MeV to zero at  $E^*\sim 12$ MeV, and an important 
quasi-free contribution, ranging from $20\%$ at $E^*\sim 3$ MeV to about $60\%$ 
at $E^*\sim 16$ MeV. This means that the total $\Delta L = 1$ contribution is 
$\sim 40 \%$ on average in the whole $E^*$ interval from 3 to 16 MeV. In 
contrast, we do not observe separate groups of events in this energy range where 
our spectrum is very smooth. As our experiment is characterized by good energy 
resolution and has excellent statistics, possible structures can not be masked 
by instrumental problems. Our data exclude any sizable contribution of $\Delta L 
= 0$ transfer and/or of uncorrelated background in this energy range.

There is a major difference between Ref.\ \cite{Yang:1995} and our work in the 
interpretation of experimental results. In Ref.\  \cite{Yang:1995} the 
structures at $E^*\sim 5.5$, 10, 15 MeV in $^{6}$Be were considered as evidence 
for important qualitative phenomena occurring in light nuclei and were 
interpreted as the low-energy ``off-springs'' of the giant-dipole and giant 
spin-dipole resonance. Our observation of the single smooth structure in the 
$E^*$ range between 3 and 15 MeV favours of the IVSDM interpretation. Though the 
IVSDM has the same transition quantum numbers as dipole and spin-dipole 
resonances, it is not a resonance but a \emph{continuum mode}, and is not a 
collective but a \emph{cluster phenomenon}.

\begin{figure}
\centerline{
\includegraphics[width=0.236\textwidth]{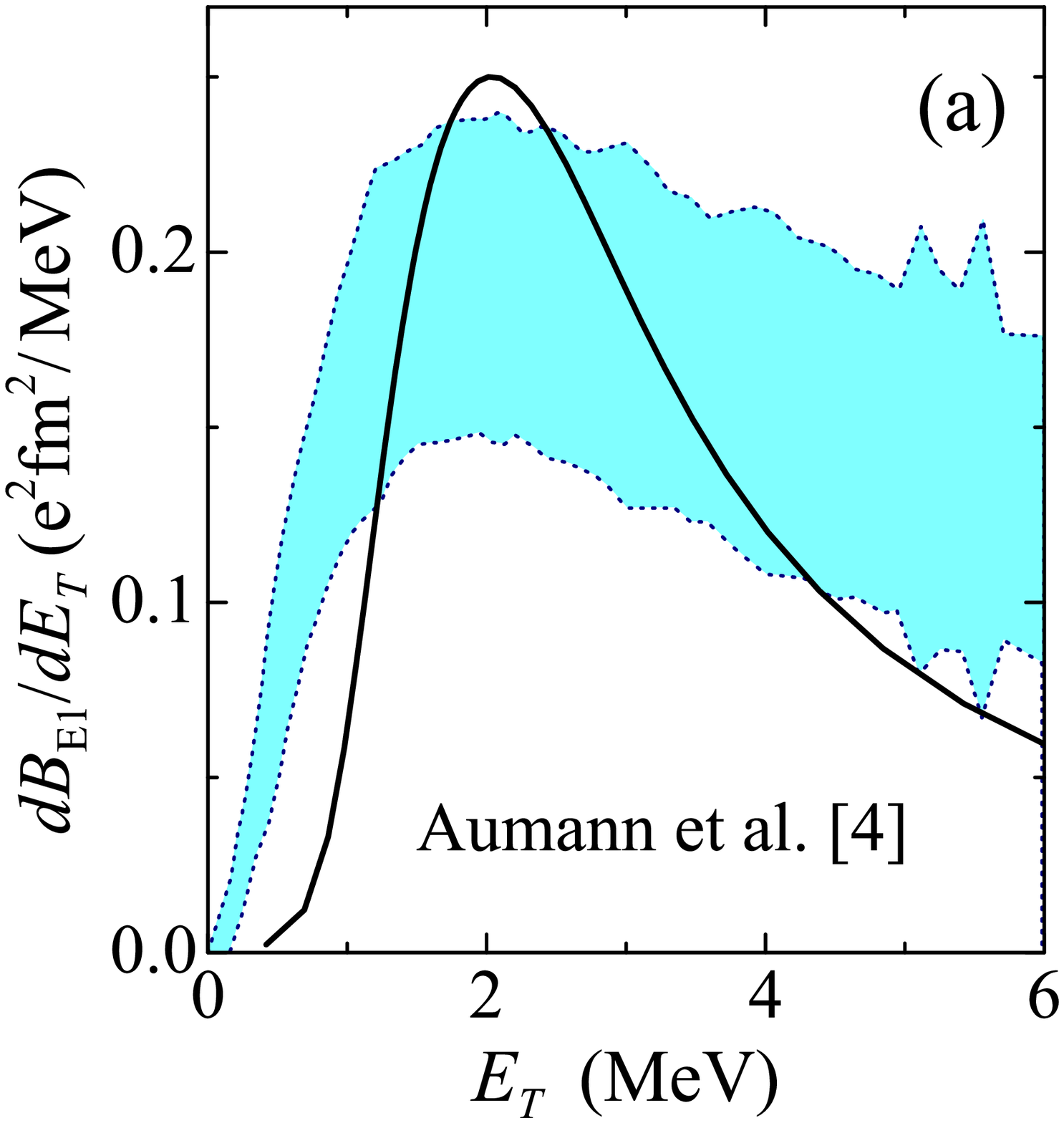}
\includegraphics[width=0.247\textwidth]{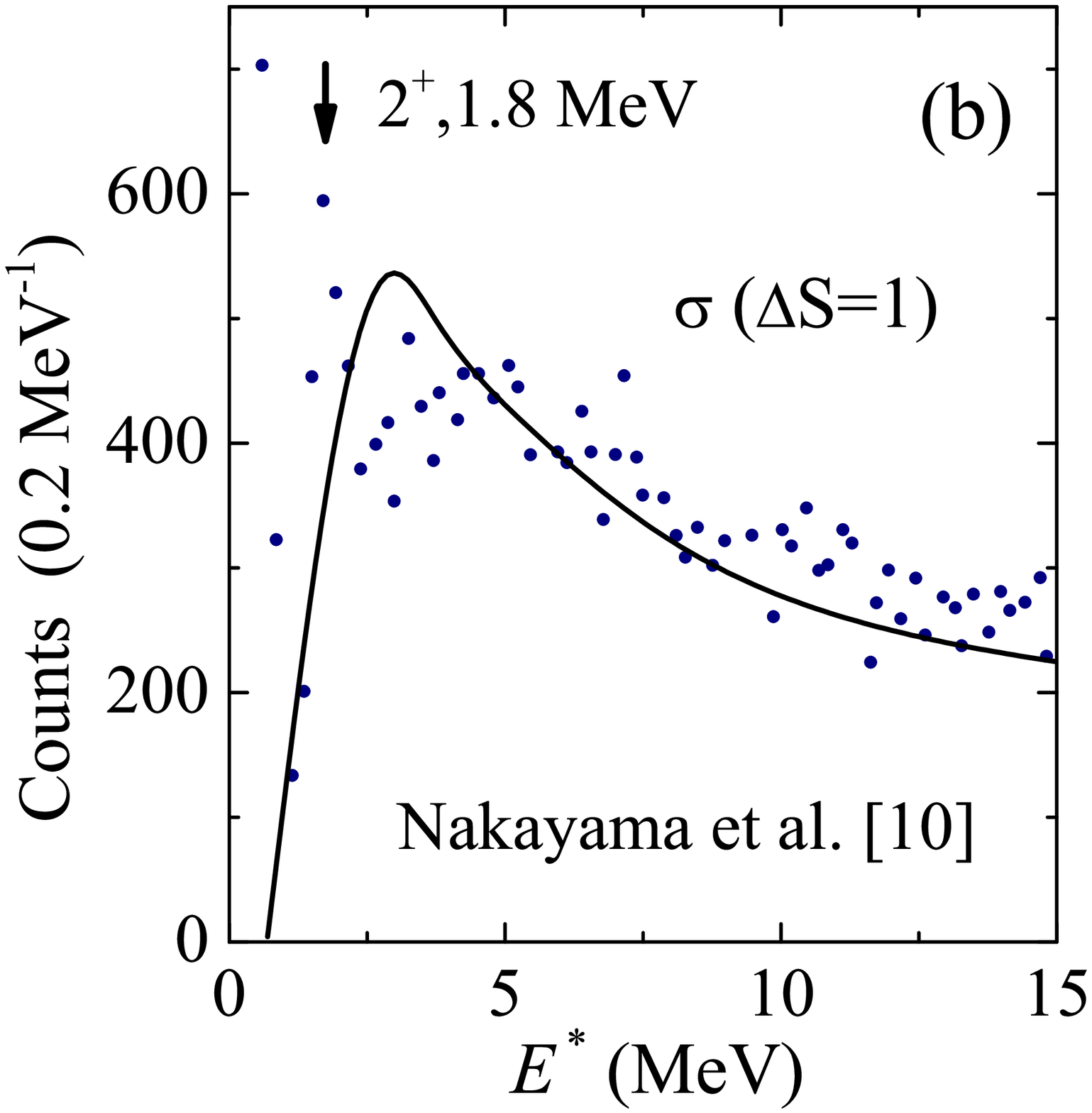}
}
\caption{Qualitative comparison (a) of SDM calculations (solid curve) with
$^{6}$He data \cite{Aumann:2005} (blue shaded range), and (b) IVSDM calculations
(solid curve) with $^{6}$He data \cite{Nakayama:2000} (diamonds).}
\label{fig:sdm-ivsdm-comp}
\end{figure}

The isovector dipole strength has been considered in the studies of ``mirror''
$^{6}$Li$(^AZ,^AZ+1)^{6}$He reactions 
\cite{Sakuta:1993,Janecke:1996,Nakayama:2000,Nakamura:2002}. The most recent 
works \cite{Nakayama:2000,Nakamura:2002} provide good-quality data and 
sophisticated interpretations. Our theoretical results indicate very similar 
behavior for the IVSDM populating the continuum of $^{6}$He and $^{6}$Be. This 
enables us to speculate about the IVSDM in $^{6}$He by using our $^{6}$Be data 
together with considerations concerning the isobaric symmetry. We suggest an 
entirely different interpretation for the $^{6}$He data obtained in Refs.\ 
\cite{Nakayama:2000,Nakamura:2002}, which is based on the following 
considerations:

(I) According to Ref.\ \cite{Nakayama:2000}, the broad structure at $E^*=4-15$ 
MeV consists of, firstly, the ``dipole excitation mode'' at $\sim 4$ MeV with 
$\Delta S=1$, secondly, the ``low-energy wing'' of the GDR at $\sim 8.5$ MeV 
formed by a $\Delta S=0$ transition, and, thirdly, the ``spin dipole resonance'' 
at $\sim 8.5$ MeV with $\Delta S=1$  which was assumed to have the same profile 
as the GDR. This work used $\gamma$-ray coincidences to distinguish the $\Delta 
S=0$, 1 contributions. However, the measurements were confined to extreme 
forward angles ($\sim 2^{\circ}$), which makes the separation of contributions 
with different $\Delta L$ a difficult task. The conclusion of Ref.\ 
\cite{Nakayama:2000} that the ``dipole excitation mode'' peak has higher energy 
than that obtained previously for SDM in Refs.\ \cite{Aumann:2005,Danilin:1998} 
can easily be explained by comparing the SDM and IVSDM spectra predicted in our 
work, see Fig.\ \ref{fig:soft-dipole}. We are unable to separate the ``dipole 
excitation mode'' and the ``spin dipole resonance'', as defined in Ref.\ 
\cite{Nakayama:2000}), in our  experiment which does not have any specific 
reaction-mechanism or quantum-number selectivity with that respect. However, our 
model calculation is able to reproduce the low-energy peak observed in the 
electromagnetic transition \cite{Aumann:2005}  [see Fig.\ 
\ref{fig:sdm-ivsdm-comp} (a)] and the broad cross-section distribution obtained 
in the charge-exchange transition \cite{Nakayama:2000} [see Fig.\ 
\ref{fig:sdm-ivsdm-comp} (b)]. Therefore, we suggest to interpret the continuum 
obtained in Ref.\ \cite{Nakayama:2000} in a range of $4 - 15$ MeV of $^{6}$He 
excitation as being due the same IVSDM phenomenon as the one proposed in the 
present work.

(II) According to Ref.\ \cite{Nakamura:2002}, two broad structures were 
populated by $\Delta L=1$ transfer in the $^{6}$Li(T,$^{3}$He)$^{6}$He reaction, 
namely one at $E^*\sim 5$ MeV, decomposed into 4.4, 7.7, and 9.9 MeV peaks, and 
another one at  $E^*\sim 15$  MeV.  Within the interpretation given in Ref.\ 
\cite{Nakamura:2002}, the underlying reaction mechanism is 
``semi-phenomenological parameterisation of quasi-free scattering'', 
contributing more than $50\%$ at $E^*\sim 10$ MeV.  The $E^*\sim 5$ MeV 
structure is interpreted as being composed of intruder states, which indicates 
quenching of the 1$p$-2$s$ shell gap in $^6$He. In our work, the correlation 
patterns displayed in Fig.\ \ref{fig:comp-mom-dis} exclude the possibility of 
any sizable phase-volume-like ``background''. The simple dynamic mechanism of 
the IVSDM population described by Eqs.\ (\ref{eq:c-e-oper}) and 
(\ref{eq:wf-plus}) does not require intruder states as the $s$-wave motion in 
the system is non-resonant, and thus does not imply any shell gap quenching.

In our experiment we obtained not only the energy profile of the $^{6}$Be 
spectrum but also the complete set of kinematic variables. We observe a 
systematic structure-less behavior of all variables with energy $E_T$ at a high 
confidence level, and do not see any reason for decomposing the continuum into 
components. We treat the continuum up to $E_T\sim 16$ MeV (at least the most 
part of the corresponding cross section) as being due to one and the same 
phenomenon, namely the  IVSDM.

Concluding this discussion, our interpretation is different from that presented 
in Refs.\ \cite{Yang:1995,Nakayama:2000,Nakamura:2002}, and can significantly 
clarify the situation concerning the ``soft'' excitations in $A = 6$ nuclei. As 
can be seen from Figs.\ \ref{fig:cross-sect} and \ref{fig:sdm-ivsdm-comp}, our 
approach provides a consistent explanation of the three different sets of data, 
namely the SDM in $^{6}$He and the IVSDM in $^{6}$He and in $^{6}$Be.

\section{Conclusion}


Accurate data on the three-body $\alpha+p+p$ continuum of $^{6}$Be system were 
obtained in the charge-exchange $p$($^{6}$Li,$^{6}$Be)$n$ reaction. The $^{6}$Be 
spectrum up to $E_T = 16$ MeV is well described by assuming the population of  
three main structures in $^{6}$Be, i.e.\ the $0^+$ state at 1.37 MeV, the $2^+$ 
state at 3.05 MeV, and a mixture of $\{0^-,1^-,2^-\}$ continuum in the $E_T$ 
range from 4 to 16 MeV. The negative-parity continuum is interpreted as a novel 
phenomenon, the IVSDM which may offer new opportunities in the nuclear structure 
studies.

%
\textit{Acknowledgments.}
%
%
--- The authors are grateful to Profs.\ Yu.Ts.\ Oganessian and S.N.\ Dmitriev 
for continuous support of this experiment. The authors are indebted to Prof.\ 
E.\ Roeckl for careful reading of the manuscript and numerous improvements. This 
work was supported by the Russian Foundation for Basic Research grant RFBR 
11-02-00657-a. L.V.G.\ is supported by FAIR-Russia Research Center grant, the  
Helmholtz  International Center for FAIR within the LOEWE program by the State 
of Hessen (Germany), and Russian Ministry of Education and Science grant 
NSh-7235.2010.2.


\bibliographystyle{elsart-num-m}
\bibliography{c:/latex/all}


\end{document}